\documentclass[prb,twocolumn,unsortedaddress,showpacs]{revtex4}

\usepackage{graphicx}

\begin{document}

\title{Influence of s-d interfacial scattering on the
magnetoresistance \\ of magnetic tunnel junctions}

\author{D.\ Bagrets}
\affiliation{
CEA/Grenoble, D\'epartement de Recherche Fondamentale sur la
Mati$\grave e$re Condens\'ee,
SP2M/NM, 38054 Grenoble, France }
\affiliation{
Department of Physics, M.~V.~Lomonosov Moscow State University,
119899 Moscow, Russia }
\author{A.\ Bagrets}
\affiliation{
Department of Physics, M.~V.~Lomonosov Moscow State University,
119899 Moscow, Russia }
\author{A.\ Vedyayev}
\affiliation{
CEA/Grenoble, D\'epartement de Recherche Fondamentale sur la
Mati$\grave e$re Condens\'ee,
SP2M/NM, 38054 Grenoble, France }
\affiliation{
Department of Physics, M.~V.~Lomonosov Moscow State University,
119899 Moscow, Russia }
\author{B.\ Dieny}
\affiliation{
CEA/Grenoble, D\'epartement de Recherche Fondamentale sur la
Mati$\grave e$re Condens\'ee,
SP2M/NM, 38054 Grenoble, France }

\date{22 August, 2001}

\begin{abstract}

We propose the two-band $s$-$d$ model to describe theoretically a diffuse
regime of the spin-dependent electron transport in magnetic tunnel junctions
(MTJ's) of the form F/O/F where F's are 3$d$ transition metal ferromagnetic 
layers and O is the insulating spacer. We aim to explain the strong 
interface sensitivity of the tunneling properties of MTJ's and investigate 
the influence of electron scattering at the nonideal interfaces on 
the degradation of the TMR magnitude. The generalized Kubo formalism and 
the Green's functions method were used to calculate the conductance of the system. 
The vertex corrections to the conductivity were found with the use of 
"ladder" approximation combined with the coherent-potential 
approximation (CPA) that allowed to consider
the case of strong electron scattering. 
It is shown that the Ward identity is satisfied in the framework of this 
approximation that provides the necessary condition for a 
conservation of a tunneling current. Based on the known results of ab-initio 
calculations of the TMR for ballistic junctions, we assume that exchange
split quasi-free\ $s$-like electrons with the density of states 
being greater for the majority spin sub-band 
give the main contribution to the TMR effect. We show that, due to 
interfacial inter-band scattering, the TMR can be substantially reduced even down 
to zero value. This is related to the fact that delocalized quasi-free electrons 
can scatter into the strongly localized $d$~sub-band with the density 
of states at the Fermi energy being larger for minority spins compared to majority 
spins. It is also shown that spin-flip electron scattering on the surface magnons
within the interface leads to a further decrease of the TMR at finite temperature.
\end{abstract}


\maketitle

\section{Introduction}

Magnetic tunnel junctions (MTJ's) consisting of two ferromagnetic layers
separated by the insulating spacer exhibit the tunneling magnetoresistance effect
(TMR) when they are switched by applying a magnetic field from
the antiparallel to parallel alignment of magnetizations in magnetic
layers. The TMR effect was first observed  by M.~Julliere in 1975 \cite{Julliere}.
Julliere found that the tunneling conductance of the trilayer structure 
Fe/Ge/Co depended on the angle between magnetizations in the Fe and Co layers. 
The measured amplitude of the TMR
in these experiments was 14\% at 4.2K. Only 20 years later, the large 
values of TMR at room temperature were obtained in magnetic junctions comprising 
the Al$_2$O$_3$ barrier \cite{Moodera,Miyazaki,Gallagher}. Since that time, there has 
been a renewed technological and fundamental interest to the tunneling phenomenon
and during the last decade a lot of experimental and theoretical 
papers were published on this topic (see reviews [5,6,7,8]). 

   Experimentally it was observed \cite{Meservey, Mathon_JMMM, Moodera_Rev} 
that the TMR depends critically on the material of the insulating barrier 
and on the conditions of its preparation, in particular on the imperfections
of the interfaces between the metal and the insulating layer \cite{JSMoodera, Mitsuzuka}. 
On the other hand, the first theory suggested by M.~Julliere expresses the TMR
ratio in terms of the effective spin polarizations $P_1$ and $P_2$
of two magnetic layers via the expression ${\rm TMR} = 2P_1 P_2/(1 + P_1 P_2)$, 
and thus predicts no dependence of the TMR on the parameters of 
the barrier. In spite of Julliere's formula is widely used for the interpretation 
of the experimental data \cite{Moodera_Rev}, it does not have rigorous theoretical 
foundation \cite{MacLaren}. The dependence of the TMR magnitude on the parameters
of the tunnel barrier and the metal\-/insulator interface was taken into account 
in the subsequent theories. Slonczewski \cite{Slonczewski} considered a quantum 
mechanical problem of tunneling of a free electron through a simple spatially uniform
barrier and showed that the TMR ratio depends on the height of the potential barrier 
and on the effective mass of the tunneling electron inside the insulator.  Later on,  
the influence of spin-flip scattering at the interfaces on temperature and 
bias-voltage dependences of the TMR was investigated by Zhang {\it et al.}\ \cite{Levy}. 
It was shown that mixing of spin-up and spin-down tunnel channels leads
to a decrease of the TMR. 

Due to a permanent progress in the development of different methods of ab-inito
calculations of the electronic properties of solids, in the last four years the
transport properties of tunnel junctions were investigated using the realistic band 
structure of the ferromagnetic layers and the insulator (Refs.~[14--20]).  
The systems where the conductance was calculated were ideal epitaxial 
Fe/\-ZnSe/\-Fe(001) \cite{MacLaren_1} and Fe/\-MgO/\-Fe(001) tunnel 
junctions \cite{Butler_1, Umerski}. For these structures the electron transport 
is assumed to be ballistic, i.e.\ the electron momentum $k_{\|}$
parallel to the ferromagnet/\-insulator interface is conserved.  
Experimentally it was also demonstrated recently by 
Heinrich {\it et al.}\ \cite{Heinrich} that Fe/\-MgO/\-Fe(001) junctions
can be indeed grown by depositing the MgO epitaxially onto a Fe whisker
and then depositing another Fe electrode epitaxially on a top of the MgO.
The results of ab-initio calculations showed that tunneling has more
complicated behavior than the predictions of the simple barrier model
proposed by Slonczewski \cite{Slonczewski}. 
The main conclusions are as follows \cite{MacLaren_1,Butler_1}. 
(i)~Tunneling conductance depends
strongly on the symmetry of the electron states in the ferromagnetic electrodes
and in the insulating layer. (ii)~The decay rates of evanescent states in the barrier
are different for the states with different symmetry. The slowest decay rates
have the evanescent states which are compatible with $s$ symmetry. The Bloch 
states in the metal couple more efficiently through the interface with the decaying
states of the same symmetry in the barrier. Therefore, mostly free-like $sp$-electrons 
from the bands with $s$ character in the ferromagnet give the essential contribution 
to the tunneling conductance. (iii) For thin insulating layers the tunneling 
current in the minority channel is dominated by the interfacial resonance states
that gives rise to "hot spots" for the $k_{\|}$-resolved conductance
in the surface Brillouin zone. The contribution from 
the resonance states is substantially 
suppressed  for the thicker barriers since the Bloch states at "hot spots" have
no $s$ character \cite{Butler_1}. Moreover, as it follows from recent
discussions \cite{Prague}, the surface resonance states are very sensitive
to the details of the interface. The asymmetry of the potential barrier
and the interfacial roughness considerably reduce the contribution 
from the surface states to a total conductance.

	Nowadays the ab-initio calculations of the TMR are possible only for  
ballistic junctions like above mentioned Fe/\-ZnSe/\-Fe and Fe/\-MgO/\-Fe
structures which are characterized by a small lattice mismatch between
the metal and the insulator, and by the well-defined band structure of the oxide or 
the semiconductor. In realistic junctions the electron transport has a diffuse
character, i.e.\ when the electron crosses the ferromagnet/insulator interface  
its $k_{\|}$ momentum is not conserved. For F/O/F 
tunnel junctions the most successful material until now has been alumina, 
Al$_2$O$_3$\cite{Moodera_Rev}. The  Al$_2$O$_3$ tunnel barriers are usually fabricated 
by the natural or plasma oxidation of the Al layer \cite{Meservey,Moodera_Rev}. 
The subsequent structural analysis,
e.g.\ with the use of the X-ray photoelectron spectroscopy, shows that
alumina is amorphous and the obtained AlO$_x$ tunnel barrier deviates 
from the ideal Al$_2$O$_3$ structure \cite{Mitsuzuka}. For a uniform 
coverage the Al film thickness is usually ranged 
from about 7 to 18~\AA,  depending on a type of the ferromagnetic 
electrode \cite{JSMoodera}.  
There is a small range of Al thicknesses that yields to the best TMR ratio 
for a given oxidation condition. When the Al layer is thin, the ferromagnet
surface becomes oxidized leading to the formation of CoO and Co$_3$O$_4$
oxides \cite{JSMoodera} or to the Fe$_3$O$_4$ oxide \cite{Mitsuzuka}. 
On the other hand, with too thick  Al film, an excess of Al metal is left 
unoxidized. The amorphous barriers, the 
roughness of the interface and its structural inhomogeneity make
the rigorous ab-initio calculations of the TMR virtually impossible
and therefore more simplified models are required to treat the diffuse 
electron transport in MTJ's.

  In our previous paper \cite{Vedyayev}, we attempted to investigate the influence
of scattering processes of the electrons (with and without spin-flip)   
at the interfaces on the TMR using a simple two-band (spin-up and spin-down)
free-electron model. It was shown that for this simplified model spin-conserving 
scattering may or may not lead to a decrease of the TMR depending on 
the amplitude of the scattering potential. In this paper, we proceed to study the 
diffuse electron transport in magnetic tunnel junctions of the from F/O/F 
with F's being 3$d$ transition metal electrodes and O being the insulating 
barrier (Al$_2$O$_3$). We use the results of ab-initio calculations to model 
the band structure of ferromagnetic electrodes, namely, the most 
important feature of these calculations that at least two groups 
of electrons form the total band structure: the almost free-like
spin-up and spin-down bands and the narrow strongly exchange split bands. 
We will call the first group $s$ and the second group $d$ electrons and 
will consider that a periodic part of the $s$-$d$ hybridization between 
bands is taken into account that results in the nonequivalence of spin-up and
spin-down $s$-bands. The parameters of the adopted model can be adjusted to 
reproduce a value of the TMR observed in the experiments. We assume that for the
case of Al$_2$O$_3$ barrier the exchange split $s$-like quasi-free 
electrons give the main contribution to the TMR effect. 
In the framework of this simplified model we will be able 
to investigate in a proper way the influence of electron 
scattering at the interfaces on the tunneling conductance. 

  To describe the nonideal tunnel junction we assume that defects and impurity 
centers (the Al or O ions, or other artificially embedded ions) are randomly 
distributed  within few monolayers near the F/O interface. 
Within the interface, an electron undergoes scattering when it 
comes to the defect or impurity center. 
We take into account these processes assuming that the hybridization between $s$ and
$d$ bands changes randomly on the interface because of this parameter 
is the most important in the adopted model among the other ones, 
characterizing the scattering potential. 
It yields to the possibility of $s$-like electrons
to scatter into the $d$ sub-band (and vice versa)
and thus strongly affects tunneling. 
To treat the electron scattering on the random potential we use the 
coherent-potential approximation (CPA) \cite{Soven} that allows to consider
the case of strong scattering. We apply the generalized Kubo formalism
and the Green's functions method to calculate the conductance of the system,
and find the vertex corrections to the conductivity with the use 
of "ladder" approximation \cite{Velicky} combined with the CPA. 
It is shown that the so-called Ward identity is satisfied in the framework 
of this approximation that provides the necessary condition for a 
conservation of a tunneling current. Note, that if it is not 
the case, the conclusions may be completely misleading. 
As a result, we show that, due to substantial difference 
in majority and minority $d$ density of states
at the Fermi energy for 3$d$ ferromagnetic metals,  
the inter-band $s$-$d$ scattering on the interface can strongly reduce the TMR 
even down to zero value. In accordance with 
results of Zhang {\it et al.}\ \cite{Levy}, 
it is also shown that the spin-flip scattering
of electrons on the surface magnons within the interface
leads to a further decrease of the 
TMR ratio at finite temperature. 

  The paper is organized as follows. In Sec.~II we describe the model 
Hamiltonian, the calculation of the tunneling conductance and vertex
corrections. The discussion of the obtained results is presented in Sec.~III.
Conclusions are in Sec.~IV. The proof of Ward identity is given 
in Appendix A. The details of the derivation of the CPA equations are described 
in Appendix B.

\section{Theoretical model}

\subsection{The Hamiltonian of the system}

 We will consider a trilayer tunnel junction of the form  
F$_1$/O/F$_3$, where F$_1$ and F$_3$ are two semi-infinite ferromagnetic
layers and O is a dielectric oxide spacer (Al$_2$O$_3$).
Our arguments in behalf of the two-band $s$-$d$ model 
which was briefly described in the introduction are as follows. 
One of the conclusions of ab-initio
calculations of the TMR for ballistic tunnel junctions \cite{MacLaren_1,Butler_1}
is that the expected spin-dependence of the tunneling current can be deduced 
from the symmetry of the Bloch states in the ferromagnet at the Fermi energy.
Spin-polarized band structure for bcc Fe, fcc Ni, and fcc Co can be found
in Ref.\ [26]. The type and symmetry of the Bloch states for different crystal 
faces with ${\bf k_{\|}} = 0$ for Fe, Co and Ni are presented in Table~I 
(in accordance with Ref.~[18]). 
For example, in case of Fe electrodes, 
the examination of the band structure shows that 
both the majority and minority bands with $s$ character in
[110] and [111] directions 
($\Lambda_1$ and $\Sigma_1$) cross the Fermi energy. 
For [100] direction the band with $s$ symmetry ($\Delta_1$) 
crosses the Fermi energy in case of the majority channel only. 
The similar analysis can also be performed in case of Ni and Co. 
Thus one can assume that in the polycrystalline Fe, Co or Ni-based films
the states with $s$ character present for both spin directions and 
the Bloch states will couple efficiently through the F/O interface with 
$s$ states in the insulator and will decay with the equal rates in the barrier
region. We will call the electrons from these bands as $s$-like electrons
and will describe them as free electrons with the effective mass 
$m_s \approx m_e$ (where $m_e$ is a bare electron mass) and with  
different Fermi momenta $k_s^{F\uparrow}$ and $k_s^{F\downarrow}$ 
($k_s^{F\uparrow} > k_s^{F\downarrow}$) for up and down spins.
The idea about a dominant contribution of the mostly itinerant
electrons to tunneling was originally proposed by M.~Stearns \cite{Stearns}
and explained the positive polarization of the spin-dependent current
in the experiments on tunneling with the superconductors \cite{Meservey}. 
According to estimations of Stearns \cite{Stearns}, in the case of Fe,
$k_{s}^{F\uparrow} = 1.09$~\AA$^{-1}$, $k_{s}^{F\downarrow} = 0.42$~\AA$^{-1}$.

\begin{table}[t]
\caption{
Type and symmetry of the Bloch states with {\bf k}$_{\|} = 0$
for Fe, Co and Ni for three different crystal faces
(in accordance with Ref.\ [18]).
The symmetry of these bands is as follows: $\Delta_1$, 
$\Sigma_1$, and $\Lambda_1$ ($s$,$p$,$d$); 
$\Delta_5$, and $\Sigma_2$ ($p$ and $d$);
and $\Delta_{2}$, $\Delta_{2'}$, $\Sigma_4$, and $\Lambda_3$ ($d$). }
\begin{ruledtabular}
\begin{tabular}{cccc}
{} & 100 & 110 & 111 \\ 
\hline
Fe$\uparrow$ & $\Delta_1$, $\Delta_{2'}$, $\Delta_5$ & 
$\Sigma_1$, $\Sigma_3$ & $\Lambda_1$ \\ 
Fe$\downarrow$ & $\Delta_2$, $\Delta_{2'}$, $\Delta_5$ & 
$\Sigma_1$, $\Sigma_3$ & $\Lambda_1$ \\ 
Co$\uparrow$ & $\Delta_1$ & $\Sigma_1$ & {} \\ 
Co$\downarrow$ & $\Delta_1$, $\Delta_5$ & $\Sigma_2$, $\Sigma_4$ & {} \\
Ni$\uparrow$ & $\Delta_1$ & $\Sigma_1$, $\Sigma_3$ & {} \\ 
Ni$\downarrow$ & $\Delta_1$, $\Delta_2$, $\Delta_5$ & 
$\Sigma_1$, $\Sigma_2$ & $\Lambda_3$ \\ \end{tabular}
\end{ruledtabular}
\end{table}

Other more localized bands (compatible with $d$ symmetry) crossing the Fermi
energy also will be described by two exchange split bands with the isotropic
quadratic dispersion law but with larger effective mass $m_d \gg m_s$.
The Fermi momenta $k_d^{F\uparrow}$ and $k_d^{F\downarrow}$ of $d$-like
electrons can be chosen to reproduce the typical for 3$d$ transition metals
ratio of the values of spin-up and spin-down $d$ density of states at the 
Fermi energy $\varepsilon_F$,
$\rho_d^{\uparrow}(\varepsilon_F) : \rho_d^{\downarrow}(\varepsilon_F)
\sim 1 : 10$ \cite{Moruzzi,Papa}.  
In accordance with the band structure of Fe, Co and Ni \cite{Moruzzi,Papa}
the narrow majority and minority $d$-bands are practically filled.
Therefore, the particles with a large effective mass, $m_d$, must be
regarded 	as holes. The values of the Fermi momenta define the positions
of band bottoms (Fig.~1) 
$V_{is}^{\mu}$, $V_{id}^{\mu}$ ($\alpha = s, d$; $i = 1,3$) 
with respect to the Fermi energy $\varepsilon_F$.
We note, that the aim of this work is to calculate the relative change of 
the TMR due to scattering. Therefore the proposed model is rather adequate 
for this purpose since the scattering
rate depends mostly on the density of states and on the matrix elements
of the scattering potential.

 The Al$_2$O$_3$ tunnel barriers obtained by oxidation of the Al film are 
amorphous \cite{Moodera_Rev,Mitsuzuka}. Concerning the $\alpha$-Al$_2$O$_3$
crystals, it is known from the band structure calculations \cite{Xu}
that the gap (which is not direct) between the upper valence band and the
conduction band is of the width of $\approx 6.29$~eV. The dispersion law 
of the lowest conduction band is not isotropic, 
and the effective electron masses along the different directions
in the Brillouin zone vary from $0.16\,m_e$ to $0.40\, m_e$ with an average
value of about $0.35\, m_e$ \cite{Xu}. This lowest conduction band is formed 
by a mixture of Al 3$s$, O 2$s$ and O 3$s$ orbitals. 
In the top of the upper valence band
the dispersion curves are very flat, i.e.\ the effective masses of holes
are large as compared with the mass of the conduction electrons. 
At the $\Gamma$ point, the top of the valence band consists of hybridized
O 2$p$ and Al 3$p$ orbitals. In a view of that, one can describe the 
conduction and valence bands of the amorphous
Al$_2$O$_3$ by isotropic quadratic 
laws with effective masses $m_s^0$ and $m_h^0$, $|m_h^0| \gg m_s^0$.

\begin{figure}[t]
\begin{center}
\includegraphics[scale=0.25, angle=-90]{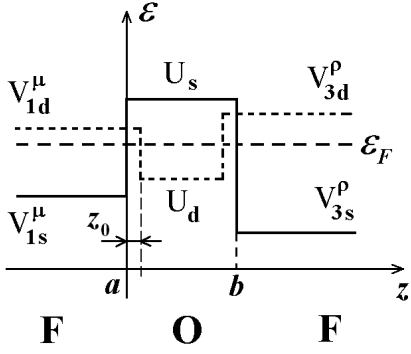}
\end{center}
\caption{The model potentials describing the propagation of
an electron in the trilayer tunnel junction F/O/F.
The solid line corresponds to the potential profile of $s$-like
electrons and the dashed line for $d$-like electrons.
$V_{1(3)\,\alpha}^{\mu(\rho)}$ denote the spin-dependent band bottoms,
$U_s$ and $U_d$ are the bottom and the top of the conduction
and valence bands in the insulator, $\varepsilon_F$ is the Fermi level, 
$z_0$ is a typical width of the interface (detail description
is given in the text).
}
\end{figure}

  Next, the $s$-like electrons from the ferromagnet can easily penetrate into the
oxide since the lead bands with $s$ character can couple efficiently with $s$ states 
in the barrier, and the tunneling conductance, caused by the specular transmission
of the Bloch waves through the interface, will decay
for both spin-up and spin-down $s$-channels with the same law as\ 
$\sim~\exp \left[ -2w \sqrt{(2m_s^0/\hbar^2)(U_s - \varepsilon_F) 
+ \kappa^2 } \right ]$, where $w$ is the width of a barrier, 
$U_s$ is a bottom of the conduction band and $\kappa = k_{\|}$ is the
electron momentum parallel to the F/O interface. Following MacLaren {\it et al.}\
\cite{MacLaren_1} we can suppose that $d$-like electrons from those bands in 
the ferromagnet without $s$ character can not couple efficiently with $s$ states 
in the oxide. Nevertheless, these $d$-bands have an admixture of $p$ symmetry
and therefore can couple with the valence bands of the Al$_2$O$_3$. However, 
in the $d$ channel the tunneling conductance due to specular transmission 
will decay very rapidly with the thickness $w$ of the barrier, as\
$\sim~\exp \left[ -2w \sqrt{(2m_h^0/\hbar^2)(\varepsilon_F - U_d) 
+ \kappa^2 } \right ]$, because of a large mass $m_h^0$ of holes in the alumina
(here $U_d$ is the top of the upper valence band). The model potentials 
describing the propagation of the electron through a tunnel junction 
are presented in Fig.~1.

  We suppose that the ferromagnet/insulator interfaces are rough and 
contaminated by impurity ions. The contamination of a few metal 
monolayers close to the F/O interface always presents after the 
oxidation of the Al film. Other impurities (e.g.\ Cr, Ru \cite{LeClair_1} or
Cu ions \cite{LeClair_2}) can be artificially inserted into the interface. 
We will characterize these structural defects by the random 
potentials, which may be divided into the spin-conserving and the spin-flip
parts.  Let $z_1 = a$ and $z_2 = b$
be the positions of interfaces, the $xy$-plane be the plane
parallel to the interface, and the $z$-axis
be the axis perpendicular to the barrier (see Fig.~1).
The Hamiltonian of the system is written as:
\begin{equation}
 \hat{H} = \hat{H}_0 + \hat{H}_{\rm spin-cons.} + \hat{H}_{\rm spin-flip},
\label{Hamilt}
\end{equation}
with
\begin{eqnarray}
 \hat{H}_0& \displaystyle = \sum_{\mu}\int d{\bf r}\, \psi^{s+}_{\mu}({\bf r})
 \left[ -\frac{\hbar^2}{2m_s(z)}\nabla^2 + U^s_{\mu}(z)
 \right] \psi^s_{\mu}({\bf r})\phantom{,}
\nonumber
\\
 {}& \displaystyle  + \sum_{\mu}\int d{\bf r}\, \psi^{d+}_{\mu}({\bf r})
 \left[ -\frac{\hbar^2}{2m_d(z)}\nabla^2 + U^d_{\mu}(z)
 \right] \psi^d_{\mu}({\bf r}),
\nonumber
\end{eqnarray}
\begin{eqnarray}
 \hat{H}_{\rm spin-cons.}& = & 
 \displaystyle \sum_{\alpha = 1,2}\sum_{\mu}\sum_{n}\int d{\bf r}
 \delta(z-z_{\alpha})\delta(\rho-\rho_n^{\alpha})\gamma_n^{\alpha}
\nonumber \\
   & \times & \left[ \psi^{s+}_{\mu}({\bf r})\psi^{d}_{\mu}({\bf r}) +
        \psi^{d+}_{\mu}({\bf r})\psi^{s}_{\mu}({\bf r})
 \right],
\nonumber
\end{eqnarray}
\begin{eqnarray}
 \hat{H}_{\rm spin-flip} =  
 \sum_{\alpha = 1,2}\sum_{n}\int d{\bf r}
 \delta(z-z_{\alpha})\delta(\rho-\rho_n^{\alpha})J_n^{\alpha} \quad \qquad
\nonumber \\  
\quad \qquad \times \left[ \psi^{s+}_{\uparrow}({\bf r})\psi^{s}_{\downarrow}({\bf r})
        \hat{S}_{-}(\rho_n^{\alpha}) +
        \psi^{s+}_{\downarrow}({\bf r})\psi^{s}_{\uparrow}({\bf r})
        \hat{S}_{+}(\rho_n^{\alpha})
 \right].
\nonumber
\end{eqnarray}
Here $\mu = \uparrow, \downarrow$ is a spin suffix;
$\psi^{s+}_{\mu}({\bf r})$, $\psi^{d+}_{\mu}({\bf r})$ and
$\psi^{s}_{\mu}({\bf r})$, $\psi^{d}_{\mu}({\bf r})$ are field operators of
the creation and annihilation of $s$ and $d$-type electrons with the spin $\mu$
at the point~${\bf r}$; 
$m_s(z)$ and $m_d(z)$ are the effective masses of electrons and holes
in the ferromagnetic layers ($m_s$ and $m_d$) or in the barrier
($m_s^0$ and $m_h^0$). $U_{\mu}^s(z)$, $U_{\mu}^d(z)$ are the
spin-dependent step-like potentials shown in Fig.~1:
\[
U_{\mu}^s(z) = \left\{
\begin{array}{ll}
V_{1s}^{\mu} & \  z<a    \\
U_{s}        & \  a<z<b  \\
V_{3s}^{\mu} & \  z>b;
\end{array}
\right.
\]
\begin{equation}
U_{\mu}^d(z) = \left\{
\begin{array}{ll}
V_{1d}^{\mu} &\  z<a + z_0    \\
U_{d}        &\  a + z_0 < z < b - z_0 \\
V_{3d}^{\mu} &\  z > b - z_0,
\end{array}
\right.
\label{U_d}
\end{equation}
where $V_{is}^{\mu}$, $V_{id}^{\mu}$ ($i = 1,3$) are the majority 
($\uparrow$) and minority ($\downarrow$) 
band bottoms in the ferromagnet, $U_s$ and $U_d$ are the 
bottom of the conduction band and the top of the upper valence band 
in the insulator. The positions of "steps" for the potential 
$U_{\mu}^d(z)$ are shifted by a value $z_0$ with respect to the 
points $z=a$ and $z=b$ in order to describe the finite thickness 
($\sim z_0$) of the interface. The explanation, why it is done 
in this way, is given below in the text.

  $\hat{H}_{\rm spin-cons.}$ is the spin-conserving part of the Hamiltonian,
$\alpha = 1,2$ are the interface numbers. 
To describe the defect structure of the nonideal F/O interface we consider
that the impurity ions and the ions of a ferromagnetic metal 
constitute the more or less random alloy of a type A$_x$B$_{1-x}$ 
where A denotes the ions of the ferromagnet (Fe, Co or Ni) and 
B denotes the impurities. Therefore, we suppose that each site $\rho_n^{\alpha}$
on the interface $\alpha$ is characterized by the random parameter
$\gamma_n^{\alpha}$ of $s$-$d$ hybridization taking 
two different values $\gamma_A$ and $\gamma_B$ with the probabilities $x$ and 
$(1-x)$, respectively. We also suppose that a periodic part of $s$-$d$
hybridization is taken into account leading to the nonequivalence of  
majority and minority $s$-bands in the ferromagnet. 

 For the simple two-band tight-binding model of the binary alloys 
\cite{Ehrenreich} one assumes that: 1) parameters 
$\varepsilon_s$ and $\varepsilon_d$ describing the positions
of $s$ and $d$ bands on the energy scale are different for 
the alloy's components; 2) also the parameter $\gamma^{A(B)}$ of the
hybridization between $s$ and $d$ bands depends on the type of an
ion (A or B). In our particular case, one can not take into account
the former effect since the well-defined two types of the electrons
will as before exist in the vicinity of the interface
and the adopted model of tunneling
will not change significantly. However, the random hybridization 
makes possible the processes of scattering of quasi-free $s$-electrons  
into the localized $d$ sub-band and vice versa 
and, therefore, can strongly influence
on tunneling.

  $\hat{H}_{\rm spin-flip}$ is a part of the Hamiltonian describing the
spin-flip scattering. We take into account only the spin-flip
processes for $s$-like electrons since these electrons 
are itinerant  and give the most 
essential contribution to the tunneling current.
Operators $\hat{S}^{+}(\rho_n^{\alpha})$, $\hat{S}^{-}(\rho_n^{\alpha})$ are 
defined as
\begin{eqnarray}
 \hat{S}_{+}(\rho_n^{\alpha}) = \frac{1}{\sqrt{2SN}}
 \sum_{\bf q}e^{iq\rho_n^{\alpha}}[ b_{\bf q} + \dots \ ],
\nonumber
\\
 \hat{S}_{-}(\rho_n^{\alpha}) = \frac{1}{\sqrt{2SN}}
 \sum_{\bf q}e^{iq\rho_n^{\alpha}}[ b^{+}_{\bf q} + \dots\ ].
\nonumber
\end{eqnarray}
Here $b^{+}_{\bf q}$ and $b_{\bf q}$ denote the creation and annihilation 
operators of the surface magnons, $N$ is a number of the lattice sites on the
interface, $S$ is a spin value. We used the well-known representation 
of the spin operators in terms of  $b_{\bf q}$ and $b^{+}_{\bf q}$ 
and left the first terms of the series. 
$J^{\alpha}_n$ is a random exchange integral which
also takes the values $J_A$ and $J_B$ 
with probabilities $x$ and $(1-x)$. 

 Let us now turn back to the step-like potential for $d$-holes. 
As it will be clear from the following consideration, 
the amplitude of the effective scattering potential on the 
interface ($z=a$) for $s$-like electrons (i.e.\ 
$ -{\rm Im}\,\Sigma^{ss}(a)$, where $\Sigma^{ss}$ is the
self-energy) is determined by the value  
$\sim -x(1-x)(\gamma_A - \gamma_B)^2 {\rm Im}\,G^{dd}(a)$,
here ${\rm Im}\,G^{dd}(a)$ is the imaginary part of the 
$d$-electron Green's function at the point $z = a$ (density of states).
If one put $z_0 = 0$ in Eq.~(\ref{U_d}), then the vertex contribution
to the conductivity (that is the contribution due to tunneling 
assisted by interfacial scattering) will be rather small as compared 
with the contribution due to direct tunneling.
This result is not enough accurate and is the sequence  
of a continual type of the model when one neglects the existence 
of the atomic lattice and for this reason 
the self-energy has a $\delta$--like
behavior on the interface.  However, the width $z_0$ of the interfacial layer 
is about the distance between atomic planes or even larger. 
For the case of bcc lattice, $z_0 = a_0/2$ for [100] direction 
(here $a_0$ is a lattice constant). 
One can show that for the present model with $z_0 = 0$
the imaginary part  $-{\rm Im}\,G^{dd}_0(z=z')$ of the unperturbed 
$d$-electron Green's function (with $\Sigma^{dd}(a) = 0$), 
i.e.\ density of states of $d$-electrons (holes),
has small value at the point of interface $z = a$
and increases inside the F-layer up to the
distance of the order of $z_0 \sim a_0$, 
and then it oscillates near the average value 
which is approximately ten times larger than
$-{\rm Im}\,G^{dd}_0(a)$.  The period of oscillations
is determined by $k_d^{F\uparrow}$ or $k_d^{F\downarrow}$, 
depending on the electron spin and on the orientation of 
magnetization in the F-layer. Such a behavior is easily understood, 
if to take into account that $d$-electrons (holes)
are almost completely reflected on the metal/\-insulator 
interface.  Thus, if one shifts the positions of "steps" 
for $U_{\mu}^d(z)$ with respect to $U_{\mu}^s(z)$ 
as it is given by (\ref{U_d}), one can expect the 
more effective mechanism of scattering due to the
larger value of $-{\rm Im}\,G^{dd}$, and therefore 
the model becomes more realistic.

\subsection{The calculation of the tunneling conductance}

To calculate the non-local conductivity we apply the Kubo formula of the linear
response \cite{Kubo} (it is valid under the small applied voltage 
which is much less than the value of the energy gap in the insulator):
\begin{eqnarray}
\sigma_{\mu\rho}({\bf r},{\bf r'})
& = & \frac{e^2}{4\pi\hbar}\ {\rm Sp}\left\{\left\langle
G_{\mu\rho}^R({\bf r},{\bf r'})
\hat D_{\bf r} \hat D_{\bf r'}
G_{\rho\mu}^A({\bf r'},{\bf r})\right\rangle \right.
\nonumber  \\
& &+\ \left. \left\langle
G_{\mu\rho}^A({\bf r},{\bf r'})
\hat D_{\bf r} \hat D_{\bf r'}
G_{\rho\mu}^R({\bf r'},{\bf r})\right\rangle
\right\},
\label{cond} 
\end{eqnarray}
where a matrix operator $\hat D_{\bf r}$ is defined as
\begin{equation}
\hat D_{\bf r} =
\left(
\begin{array}{cc}
\displaystyle \frac{1}{2m_s(z)}\stackrel{\leftrightarrow}{\nabla}_{\bf r} & 0 \\
0 & \displaystyle \frac{1}{2m_d(z)}\stackrel{\leftrightarrow}{\nabla}_{\bf r}
\end{array}
\right),
\label{D_r}
\end{equation}
and $\stackrel{\leftrightarrow}{\nabla}_{\bf r} =
(\stackrel{\rightarrow}{\nabla}_{\bf r} - \stackrel{\leftarrow}{\nabla}_{\bf r})$ is
the asymmetric gradient operator, 
$G_{\mu\rho}^R({\bf r},{\bf r'})$ and $G_{\mu\rho}^A({\bf r},{\bf r'})$
are the retarded and advanced $(2\times 2)$--matrix Green's functions
(with components $ss$, $sd$, $ds$ and $dd$),
$\mu,\rho = \uparrow, \downarrow$ are the
spin suffixes, brackets $\langle\dots\rangle$ denote the averaging over the
configurations and magnon degrees of freedom, the trace (Sp) goes 
over $s$ and $d$ indices of the bands.
Below, for convenience, it is assumed that $\hbar = 1$. We will recall
about $\hbar$ in the final expressions for the conductance.

  To calculate the conductivity (\ref{cond}) of the system one has to find
the Green's function of the Hamiltonian (\ref{Hamilt}), which can be found by solving
the following system of differential equations in the mixed
$(\kappa,z)$ representation \cite{Vedy}:
\begin{equation}
\sum_{\gamma = s,d}
\left[\varepsilon\, \delta^{\alpha\gamma}
- \hat H_{\mu}^{\alpha\gamma}(z)\right]G_{\mu\kappa}^{\gamma\beta}(z,z')
 = \delta_{\alpha\beta}\delta(z-z'),
\label{Green_eq} 
\end{equation}
\begin{eqnarray}
\label{H_eff}
\hat H^{\alpha\gamma}_{\mu}(z) & = & \displaystyle
\left\{ - \frac{1}{2 m_{\alpha}}\frac{\partial^2}{\partial z^2}  +
\frac{\kappa^2}{2 m_{\alpha}} + U_{\mu}^{\alpha}(z) \right\} \delta^{\alpha\gamma} \\
& + & \displaystyle \Sigma_{\mu}^{\alpha\gamma}(a)\delta(z-a) \, + \,
\Sigma_{\mu}^{\alpha\gamma}(b)\delta(z-b),
\nonumber
\end{eqnarray}
where 
$\kappa = k_{\|}$ is a projection of the electron momentum on the $xy$--plane
(parallel to the F/O interface),
the Greek indices $\alpha$, $\beta$ and $\gamma$ assume values $s$ and $d$,
$\mu$ denotes the electron spin.
$\hat H^{\alpha\beta}_{\mu}(z)$ is the $(2\times 2)$--matrix linear
differential operator  where $\Sigma_{\mu}^{\alpha\beta}(a)$ and
$\Sigma_{\mu}^{\alpha\beta}(b)$
($\alpha,\beta = s,d$; $\mu = \uparrow, \downarrow$)
denote the coherent potentials for spin-up and spin-down electrons
which take into account the scattering of the electron
by random spin-conserving and spin-flip
potentials on the interfaces. They were found with the use of 
the coherent potential approximation (CPA) \cite{Soven}, the details of
these calculations are presented in the subsequent section~IIC 
and in Appendix B. The operator $\hat H_{\mu}(z)$ represents the 
effective single-particle Hamiltonian of the system which, however, is
non-Hermitian since coherent potentials are imaginary quantities.

  In order to solve Eq.~(\ref{Green_eq}) 
for the Green's functions, we will follow the
procedure described below. First, let's solve the Shr\"{o}dinger equation
with the Hamiltonian $\hat H_{\mu}(z)$:
\begin{equation}
\sum_{\beta = s,d} \left[\varepsilon\, \delta^{\alpha\beta}
- \hat H^{\alpha\beta}_{\mu}(z)\right]
\psi_{\beta}(z) = 0.
\label{Schrod}
\end{equation}
The solutions of this equation can be easily found
since the potentials $U^{\alpha}_{\mu}(z)$
have a step-like form. Let us put $\varepsilon = \varepsilon_F + i0$, 
where $\varepsilon_F$ is the Fermi energy,
and introduce the following notations:
\[
k^{F\mu}_{1s} = \sqrt{2m_s(\varepsilon_F - V^{\mu}_{1s})}, 
\quad
k^{F\mu}_{3s} = \sqrt{2m_s(\varepsilon_F - V^{\mu}_{3s})},
\]
\[
k^{F\mu}_{1d} = \sqrt{2m_d(\varepsilon_F - V^{\mu}_{1d})},
\quad
k^{F\mu}_{3d} = \sqrt{2m_d(\varepsilon_F - V^{\mu}_{3d})}
\]
are Fermi momenta in F$_1$ and F$_3$ ferromagnetic layers
for electrons with the spin $\mu$, and
\[
q^{0}_{2s} = \sqrt{2m^0_s(U_s - \varepsilon_F)}, \quad 
q^{0}_{2d} = \sqrt{2m^0_h(U_d - \varepsilon_F)}. 
\]
Let also
$k_1^{\mu\alpha} = \sqrt{ (k_{1\alpha}^{F\mu})^2 - \kappa^2}$ and
$k_3^{\mu\alpha} = \sqrt{ (k_{3\alpha}^{F\mu})^2 - \kappa^2}$
be the components of electron momentum with spin $\mu$ along $z$-axis in
F$_1$ and F$_3$ layers, respectively
($\kappa$ is the in-plane component of the momentum,
$\alpha$ is a band index), and let 
$q_2^{\alpha} = \sqrt{(q_{2\alpha}^{0})^2 + \kappa^2}$
be the imaginary electron momentum in the insulating layer.

Further, for convenience, we will omit indices $\mu$ and $\kappa$ in
the notation of some functions. Equation (\ref{Schrod}) has four
linear-independent solutions which we denote as
\[
\psi_i(z) = \left(
\begin{array}{c} \psi_i^s(z) \\ \psi_i^d(z) \end{array}
\right) \quad (i = 1,2)
\]
and
\[
\varphi_i(z) = \left(
\begin{array}{c} \varphi_i^s(z) \\ \varphi_i^d(z) \end{array}
\right) \quad (i = 1,2).
\]
We choose these independent solutions so that
two functions $\psi_1(z)$ and $\varphi_1(z)$
would describe two waves corresponding to the $s$-like electrons,
and functions $\psi_2(z)$ and $\varphi_2(z)$ would correspond to
the $d$-like electrons. Namely, in a layer F$_1$
$(z<a)$ the solutions $\varphi_i(z)$ have
the form
\begin{equation}
\varphi_1(z) = \left(
\begin{array}{c} \exp[-ik_1^{\mu s}z] \\ 0 \end{array}
\right) \quad z<a,
\label{phi}
\end{equation}
\[
\varphi_2(z) = \left(
\begin{array}{c} 0 \\ \exp[-ik_1^{\mu d}z] \end{array}
\right) \quad z<a,
\]
and in a layer F$_3$ $(z>b)$ the solutions $\psi_i(z)$ are
\begin{equation}
\psi_1(z) = \left(
\begin{array}{c} \exp[ik_3^{\mu s}z] \\ 0 \end{array}
\right) \quad z>b,
\label{psi}
\end{equation}
\[
\psi_2(z) = \left(
\begin{array}{c} 0 \\ \exp[ik_3^{\mu d}z] \end{array}
\right) \quad z>b.
\]
Since $\varepsilon = \varepsilon_F + i0$, then
 ${\rm Im}\,k_i^{\mu\alpha} = +0$
($i=1,2$; $\alpha = s,d$).
Thus these solutions satisfy the following boundary conditions:
\begin{eqnarray}
\psi_i(z) \to 0 & \quad \mbox{if}\quad z\to +\infty
\quad (i = 1,2),
\nonumber
\\
\varphi_i(z) \to 0 & \quad \mbox{if}\quad z\to -\infty
\quad (i = 1,2).
\nonumber
\end{eqnarray}
Starting from expressions (\ref{phi}) and (\ref{psi}),
the solutions $\varphi_i(z)$, $\psi_i(z)$ can be easily extended
in two other layers. Let us introduce the matrices
\[
\Phi(z) =
\left(
\begin{array}{cc}
\varphi_1^s(z) & \varphi_2^s(z) \\
\varphi_1^d(z) & \varphi_2^d(z)
\end{array}
\right),
\]
\[
\Psi(z) =
\left(
\begin{array}{cc}
\psi_1^s(z) & \psi_2^s(z) \\
\psi_1^d(z) & \psi_2^d(z)
\end{array}
\right).
\]
The Wronskian of the system (\ref{Schrod}) is
\begin{equation}
\Delta = \Phi^T(z)\hat D_z \Psi(z).
\label{Wronsc}
\end{equation}
where the matrix operator $\hat D_z$
is defined similar to Eq.\ (\ref{D_r}), 
and the subscript $T$ denotes the transposition operation.
It is known from the theory of differential equations that
the matrix $\Delta$ is a constant since it satisfies the equation
$\displaystyle \frac{\partial\Delta(z)}{\partial z} = 0$.
Taking into account that $\varepsilon = \varepsilon_F + i0$ and
${\rm Im}\ \Sigma^{\alpha\beta}_{\mu} < 0$,
the solution of Eq.~(\ref{Green_eq}) for the retarded Green's function
can be written in the matrix form as
\begin{eqnarray}
G^R(z,z') & = & \Phi(z)[\Delta^T]^{-1}\Psi^T(z'), 
\quad \mbox{if}\quad z<z',
\label{Green_R}
\\
G^R(z,z') & = & \Psi(z)\Delta^{-1}\Phi^T(z'), 
\qquad \mbox{if}\quad z>z'.
\nonumber
\end{eqnarray}
To find the advanced Green's function, one has to put 
$\varepsilon = \varepsilon_F - i0$
in Eq.~(\ref{Green_eq}) and assume
that ${\rm Im}\ \Sigma^{\alpha\beta}_{\mu} > 0$.
Then we obtain
\begin{equation}
G^A(z,z') = \left[G^R(z,z')\right]^{*}.
\label{Green_A}
\end{equation}
In the expressions (\ref{Green_R}) and (\ref{Green_A})
the Green's functions depend on the in-plane momentum $\kappa$
and on the spin $\mu$ of the electron because solutions
$\psi_i(z)$ and $\varphi_i(z)$ also depend on $\mu$ and $\kappa$.

 Next, one has to find the two-point conductivity (\ref{cond})
of the system using the Green's functions (\ref{Green_R})
and (\ref{Green_A}). For that, we introduce the current matrices
$j_{\mu}^{\psi}$ and $j_{\mu}^{\varphi}$ ($\mu = \uparrow, \downarrow$)
constructed with the use of solutions $\psi_i(z)$ and $\varphi_i(z)$, respectively:
\begin{eqnarray}
j^{\psi}(z) & = & -i \Psi^{\dagger}(z)\hat D_z \Psi(z),
\label{j_current}
\\
j^{\varphi}(z) & = & -i \Phi^{\dagger}(z)\hat D_z \Phi(z).
\nonumber
\end{eqnarray}
The total conductance of the system $\sigma_{\mu\rho}(z,z')$ may be
presented in the usual form as a sum of the "bubble" part and the 
vertex corrections (see Fig.~2) \cite{Levy2}:
\begin{equation}
\sigma_{\mu\rho}(z,z') = \delta_{\mu\rho}\sigma^0_{\mu}(z,z') +
\sigma^{\Gamma_a}_{\mu\rho}(z,z') + \sigma^{\Gamma_b}_{\mu\rho}(z,z').
\end{equation}

The "bubble" contribution to the conductance describes
direct electron tunneling from the electrode to another one 
through the barrier when electron momentum
parallel to the F/O interface $\kappa = k_{\|}$ 
is conserved (the specular transmission). 
The scattering on the interfaces affects direct
tunneling so that the effective height of the potential barrier
increases because the electron also has to pass through 
$\delta$-like potentials at the points $z = a$ and $z = b$ formed
by the self-energies $\Sigma_{\mu}(a)$ and $\Sigma_{\mu}(b)$.
The vertex corrections to the conductance 
describe the tunneling assisted by interfacial
roughness --- that is the processes 
when the electron with momentum $\kappa$ 
comes to the impurity center on the interface, 
undergoes scattering ($\kappa \to \kappa'$) in another
channel with $\kappa' \ne \kappa$, and then goes away 
to the electrode or to the barrier.

It also can be shown that the contribution to the tunneling 
conductance is negligibly small ($\sim e^{-4q_2^s (b-a) }$) 
from the diagram containing both vertex parts $\Gamma_a$ and 
$\Gamma_b$ compared with other contributions that are of the order 
of $\sim e^{-2q_2^s (b-a)}$. The diagram with vertices 
$\Gamma_a$ and $\Gamma_b$ corresponds to the interference  
of waves scattered from both interfaces. Therefore, we
can neglect this interference term.

\begin{figure}[t]
\begin{center}
\includegraphics[scale=0.35, angle=0]{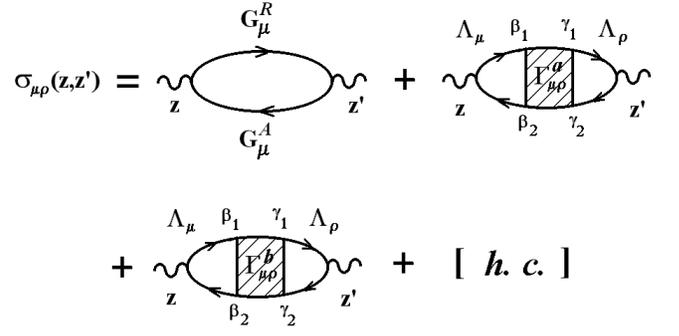}
\end{center}
\caption{The diagrammatic representation of the total two-point nonlocal
conductivity $\sigma_{\mu\rho}(z,z')$ as a sum
of "bubble" and "vertex" parts. Here the full lines correspond to 
the Green's functions $G^{R}(z,z')$ and $G^{A}(z,z')$, and
wavy lines denote the asymmetric gradient operator
$\stackrel{\leftrightarrow}{\nabla}_z$
of velocity at the points $z$ and $z'$. The shaded square designates
the vertex part $\Gamma^{\mu\rho}$ at the interface;
{\it h.c.}\ denotes the complex conjugate terms.}
\end{figure}

Substituting obtained expressions (\ref{Green_R}), (\ref{Green_A})
for the Green's functions and using the definition (\ref{j_current})
for the current matrices, we come to the following results.
The analytical expression for the "bubble" conductance is given by
\begin{equation}
\sigma^0_{\mu}(z,z') = -\frac{e^2}{2\pi\hbar A}\sum_{\kappa}{\rm Sp}
\left\{
\frac{1}{\Delta_{\mu}}
\left[j_{\mu}^{\varphi}(z)\right]^T\frac{1}{\Delta_{\mu}^{\dagger}}
j_{\mu}^{\psi}(z')\right\},
\label{s_0}
\end{equation}
here it is assumed that $z < z'$, $\Delta_{\mu}$ is the Wronskian 
(\ref{Wronsc}), $A$ denotes the junction area.

 For $z < a$, $z' > b$, the vertex corrections from the left and 
from the right interfaces can be written as
\begin{eqnarray}
\sigma^{\Gamma_a}_{\mu\rho}(z,z') & = &
-\frac{e^2}{2\pi\hbar A^2}\sum_{\kappa\kappa'}{\rm Sp}
\Bigl\{
\Lambda^{\beta_1\beta_2}_{\mu\kappa}(z,a) \nonumber  \\
& & \times\ \Gamma^{\mu\rho}_a
 \left({\scriptsize \begin{array}{cc}
 \beta_1 & \gamma_1\\
 \beta_2 & \gamma_2
 \end{array} }\right)
\Lambda^{\gamma_1\gamma_2}_{\rho\kappa'}(a,z')
\Bigr\}
\nonumber \\
\sigma^{\Gamma_b}_{\mu\rho}(z,z') & = &
-\frac{e^2}{2\pi\hbar A^2}\sum_{\kappa\kappa'}{\rm Sp}
\Bigl\{
\Lambda^{\beta_1\beta_2}_{\mu\kappa}(z,b) 
\label{s_vertex} \\
& & \times\ 
\Gamma^{\mu\rho}_b
 \left({\scriptsize \begin{array}{cc}
 \beta_1 & \gamma_1\\
 \beta_2 & \gamma_2
 \end{array} }\right)
\Lambda^{\gamma_1\gamma_2}_{\rho\kappa'}(b,z')
\Bigr\} \nonumber
\end{eqnarray}
where summation is also performed over repeating
indices $\beta_i, \gamma_i = s,d$. Here
$\Gamma^{\mu\rho}_a
 \left({\scriptsize \begin{array}{cc}
 \beta_1 & \gamma_1\\
 \beta_2 & \gamma_2
 \end{array} }\right)
$
and
$
\Gamma^{\mu\rho}_b
 \left({\scriptsize \begin{array}{cc}
 \beta_1 & \gamma_1\\
 \beta_2 & \gamma_2
 \end{array} }\right)
$
are the vertex parts on the interfaces $a$ and $b$.
$\Lambda_{\mu\kappa}$ are $(2\times 2)$--matrices, their components
are defined by
\begin{eqnarray}
\Lambda^{\beta_1\beta_2}_{\mu\kappa}(z,a) = {\rm Sp}
\left\{
\frac{1}{\Delta_{\mu}}
j_{\mu}^{\varphi T}(z) \frac{1}{\Delta_{\mu}^{\dagger}}
\left[\rho^{\psi}_{\mu}(a)\right]^{\beta_1\beta_2}
\right\},
\label{Lambda}
\\
\Lambda^{\gamma_1\gamma_2}_{\mu\kappa}(a,z') = {\rm Sp}
\left\{
\frac{1}{\Delta_{\mu}}
\left[\rho^{\varphi T}_{\mu}(a)\right]^{\gamma_1\gamma_2}
\frac{1}{\Delta_{\mu}^{\dagger}}
j^{\psi}(z')
\right\},
\nonumber
\end{eqnarray}
where $\left[\rho^{\psi}_{\mu}(a)\right]^{\beta_1\beta_2}$ and
$\left[\rho^{\varphi}_{\mu}(a)\right]^{\gamma_1\gamma_2}$ are
density matrices with the components:
\begin{eqnarray}
\left[\rho^{\psi}_{\mu}(a)\right]^{\beta_1\beta_2}_{ik} & = &
\psi^{\beta_1*}_i(a)\psi^{\beta_2}_k(a)\quad (i,k = s,d),
\label{rho}
\\
\left[\rho^{\varphi}_{\mu}(a)\right]^{\gamma_1\gamma_2}_{ik} & = &
\varphi^{\gamma_1*}_i(a)\varphi^{\gamma_2}_k(a)\quad (i,k = s,d).
\nonumber
\end{eqnarray}
The expressions similar to expressions (\ref{Lambda}) may be also
written for the matrices $\Lambda_{\mu\kappa}(z,b)$,
$\Lambda_{\rho\kappa}(b,z')$ and for 
other position of the points $z$, $z'$
with respect to $a$ and $b$.
The vertex parts $\Gamma^{\mu\rho}_a$ and
$\Gamma^{\mu\rho}_b$ are calculated in the
"ladder" approximation \cite{Velicky}. The derivation
of the equation for $\Gamma^{\mu\rho}$ is presented below 
in the subsection IID.

  We have to note, that coherent potentials $\Sigma_{\mu}(a)$,
$\Sigma_{\mu}(b)$ calculated in the framework of the CPA and vertices
$\Gamma^{\mu\rho}_a$, $\Gamma^{\mu\rho}_b$ calculated in the "ladder"
approximation satisfy the so-called Ward identity 
which in our case, for example, for the interface $z=a$ can be 
written as follows (for the details, see Appendix A):
\begin{eqnarray}
{\rm Im}\, \Sigma_{\mu}^{\beta_1\beta_2}(a) & = & \sum_{\rho=\uparrow,\downarrow}
\Gamma_a^{\mu\rho}
 \left({\scriptsize \begin{array}{cc}
 \beta_1 & \gamma_1 \\
 \beta_2 & \gamma_2
 \end{array} }\right)
 \frac{1}{A}\sum_{\kappa}
 \Bigl\{ {\rm Im}\, G_{\rho\kappa}^{\gamma_1\gamma_2}(a)
\nonumber
\\
 & - & G_{\rho\kappa}^{\gamma_1\alpha_1*}(a)\,
   G_{\rho\kappa}^{\gamma_2\alpha_2}(a)\,
   {\rm Im}\, \Sigma_{\rho\kappa}^{\alpha_1\alpha_2}(a)
 \Bigr\}.\quad 
\label{Ward}
\end{eqnarray}
Here the summation is also performed over repeating indices 
$\gamma_i$  and $\alpha_i$. The fulfillment of (\ref{Ward}) provides 
the necessary condition of the nondivergence of the current through
the tunnel junction:
\begin{equation}
\frac{\partial}{\partial z'}\sigma^0_{\mu}(z,z') +
\frac{\partial}{\partial z'}
\sum_{\rho = \uparrow, \downarrow}
\left[
\sigma_{\mu\rho}^{\Gamma_a}(z,z') +
\sigma_{\mu\rho}^{\Gamma_b}(z,z')
\right] = 0.
\label{current0}
\end{equation}

   According to Eq.~(\ref{current0}), the total conductance of the
system is a constant value. In view of that, we will derive the exact expression 
for $\sigma_{\mu\rho}(z,z')$ evaluating the conductance at points $z = a-0$ and
$z'=b+0$, i.e.\ at the left and at the right sides from the interface.
The Wronskian matrix $\Delta_{\mu}$ and the current matrices $j^{\varphi}_{\mu}(z)$
and $j^{\psi}_{\mu}(z')$ are expressed in terms of matrices
$\Phi(z)$ and $\Psi(z)$ --- these matrices, as follows from Eq.~(\ref{Green_R}),
determine the Green's function. The straightforward evaluation of the
conductance according to formulae (\ref{s_0}--\ref{Lambda}) leads to
the result that $\sigma_{\mu\rho}(a,b)$ is expressed in terms
of the retarded Green's functions at the points of interfaces 
$G_{\mu}(a) = G^R_{\mu}(z=z'=a)$ and
$G_{\mu}(b) = G^R_{\mu}(z=z'=b)$ constructed
according to Eq.~(\ref{Green_R}). The explicit form 
of $G_{\mu}(a)$ in the $(\kappa,z)$ representation is given by the
expression:
\begin{widetext}
\begin{equation}
 G_{\mu}(a)  =  \left(
 \begin{array}{cc}
 \displaystyle
 \frac{k_1^{\mu\,s}}{2m_s}(i - {\rm cotan}\,\varphi_1^{\mu\,s}) -
 \Sigma_{\mu}^{ss}(a) &  - \Sigma_{\mu}^{sd}(a) \\
 - \Sigma_{\mu}^{ds}(a) &
 \displaystyle
 \frac{k_1^{\mu\,d}}{2m_d}\left[i - {\rm cotan}\,
 (k_1^d z_0+\varphi_1^{\mu\,d})\right] 
 - \Sigma_{\mu}^{dd}(a)
\end{array}\right)^{-1},
\label{G_int}
\end{equation}
\end{widetext}
where
\[
{\rm tan}\,\varphi_1^{\mu\,s} = \frac{k_1^{\mu\,s}}{q_2^s}\frac{m_s^0}{m_s},
\quad
{\rm tan}\,\varphi_1^{\mu\,d} = \frac{k_1^{\mu\,d}}{q_2^d}\frac{m_h^0}{m_d}.
\]
The similar expression can be written for the Green's function 
$G_{\mu}(b)$ on the right interface. For that, one 
has to make the substitutions
$k_1^{\mu\,\alpha} \to k_3^{\mu\,\alpha}$, 
$\Sigma_{\mu}^{\alpha\beta}(a) \to \Sigma_{\mu}^{\alpha\beta}(b)$,
$\varphi_1^{\mu\, \alpha} \to \varphi_3^{\mu\, \alpha}$
in the expression for $G_{\mu}(a)$.
Here $m_{\alpha}$ ($\alpha = s,d$) and $m^0_{s(h)}$ are the effective
electron (hole) masses in the ferromagnetic and insulating layers;
$k_i^{\mu\alpha}$ and $q_2^{\alpha}$ ($i=1,3$, $\alpha = s,d$) are the
functions on $\kappa$ introduced above in the text after Eq.~(\ref{Schrod}).
Let us define the "transport" density of states as follows:
\begin{eqnarray}
 A_{\mu}(a) & = & G_{\mu}^{\dagger}(a)\tilde{j}_{\mu}^{\varphi}G_{\mu}(a),
\nonumber
\\
 A_{\mu}(b) & = & G_{\mu}^{\dagger}(b)\tilde{j}_{\mu}^{\psi}G_{\mu}(b),
\label{Trans}
\end{eqnarray}
where
\[
\tilde{j}_{\mu}^{\varphi} = \left(
\begin{array}{cc}
\displaystyle
\frac{k_1^{\mu\,s}}{m_s} & 0 \\
0 &\displaystyle \frac{k_1^{\mu\,d}}{m_d}
\end{array}\right), \qquad
\tilde{j}_{\mu}^{\psi} = \left(
\begin{array}{cc}
\displaystyle
\frac{k_3^{\mu\,s}}{m_s} & 0 \\
0 &\displaystyle \frac{k_3^{\mu\,d}}{m_d}
\end{array}\right).
\]

  Expression (\ref{s_0}) for the "bubble" conductance then reads
\begin{eqnarray}
\sigma^{0}_{\mu}(a,b) & = & \frac{e^2}{2\pi\hbar A}\sum_{\kappa}{\rm Sp}\left\{
\lambda_b^{-1}\hat{\left(\frac{q}{m}\right)}\lambda_a^{-1} \right.
\nonumber \\
& & \times \left. A^T_{\mu}(a)\lambda_a^{-1}\hat{\left(\frac{q}{m}\right)}
\lambda_b^{-1} A_{\mu}(b) \right\},
\label{s_0_ab}
\end{eqnarray}
where
\[
\lambda_a = \left(
\begin{array}{cc}
e^{q_2^s z_0} & 0 \\
0 & \lambda_a^d
\end{array}\right), \qquad
\lambda_a^d = \frac{\sin(k_1^{\mu\, d} z_0 + \varphi_1^{\mu\, d})}
 {\sin\varphi_1^{\mu\, d}},
\]
\[
\lambda_b = \left(
\begin{array}{cc}
e^{q_2^s z_0} & 0 \\
0 & \lambda_b^d
\end{array}\right), \qquad
\lambda_b^d = \frac{\sin(k_3^{\mu\, d}z_0 + 
 \varphi_3^{\mu\, d})}{\sin\varphi_3^{\mu\, d}},
\]
\[
 \hat{\left(\frac{q}{m}\right)} =
 \left(
 \begin{array}{cc}
   \displaystyle
   \frac{q^s_2}{m_s^0} e^{-q_2^s\tilde w} & 0 \\
   0  & \displaystyle \frac{q^d_2}{m_h^0} e^{-q_2^d \tilde w}
 \end{array}\right),
\]
and $\tilde w = b - a - 2z_0$ is the "width" of $d$--barrier.
 
For the vertex correction from the left 
interface ($z=a$) we obtain
\begin{eqnarray}
 \sigma^{\Gamma_a}_{\mu\rho}(a,b) & = & -\frac{e^2}
 {2\pi\hbar A^2}\sum_{\kappa\kappa'}
 \Lambda^{\beta_1\beta_2}_{\mu\,\kappa}(a,a) \nonumber \\
 & & \times\ \Gamma_a^{\mu\rho}
 \left({\scriptsize \begin{array}{cc}
 \beta_1 & \gamma_1 \\
 \beta_2 & \gamma_2
 \end{array} }\right)\Lambda^{\gamma_1\gamma_2}_{\rho\,\kappa'}(a,b),
\label{s_vertex_a} \\
\Lambda_{\mu}(a,a) & = & - A_{\mu}(a), \nonumber \\ 
\Lambda_{\rho}(a,b) & = & G_{\rho}^{\dagger}(a)\lambda_a^{-1}
\hat{\left(\frac{q}{m}\right)}\lambda_b^{-1} A_{\rho}(b)
\nonumber \\ 
& & \times\ \lambda_b^{-1}\hat{\left(\frac{q}{m}\right)}
\lambda_a^{-1}G_{\rho}(a).
\nonumber
\end{eqnarray}
\noindent
In the similar way, the vertex correction from 
the right interface ($z=b$) reads 
\begin{eqnarray}
 \sigma^{\Gamma_b}_{\mu\rho}(a,b) & = & -\frac{e^2}
 {2\pi\hbar A^2}\sum_{\kappa\kappa'} 
 \Lambda^{\beta_1\beta_2}_{\mu\,\kappa}(a,b) \nonumber \\
& & \times\ \Gamma_b^{\mu\rho}
 \left({\scriptsize \begin{array}{cc}
 \beta_1 & \gamma_1\\
 \beta_2 & \gamma_2
 \end{array} }\right)\Lambda^{\gamma_1\gamma_2}_{\rho\,\kappa'}(b,b),
\label{s_vertex_b} \\
 \Lambda_{\mu}(a,b) & = & G^{\dagger}_{\mu}(b)\lambda_b^{-1}
 \hat{\left(\frac{q}{m}\right)}\lambda_a^{-1}  A_{\mu}(a) 
\nonumber \\
& & \times\ \lambda_a^{-1}\hat{\left(\frac{q}{m}\right)}
 \lambda_b^{-1}G_{\mu}(b), \nonumber \\
 \Lambda_{\rho}(b,b)& = & - A_{\rho}(b).
\quad \nonumber
\end{eqnarray}
Expressions (\ref{s_0_ab}--\ref{s_vertex_b}) determine the total conductance
of the system.

\subsection{The CPA equations}

The scattering of the electron by random potentials within the interface
(terms $\hat H_{\rm spin-cons.}$ and $\hat H_{\rm spin-flip}$ in the 
Hamiltonian) is taken into account by introducing the self-energy operators
$\Sigma_{\mu}^{\alpha\beta}(a)$ and $\Sigma_{\mu}^{\alpha\beta}(b)$ into the
effective single-particle Hamiltionian (\ref{H_eff}). To calculate the
self-energies we apply the coherent-potential approximation (CPA) \cite{Soven,Ehrenreich}.
Let's denote kets $|\gamma,\mu,\rho_n^{\alpha}\rangle =
|\gamma, \mu \rangle\otimes |\rho_n^{\alpha}\rangle$
corresponding to the Wannier states of
the electron on the interface $\alpha$ ($\alpha=1,2$)
at the given site $\rho_n^{\alpha}$ in the $(x,y)$-plane,
where $\gamma$ refers to the state $s$ or $d$ and $\mu$ is the spin.  
The symbol $\otimes$ denotes the direct product. 
The problem of finding of the single-particle Green's 
function $G_{\mu\rho}({\bf r}, {\bf r'})$ of the
many-body Hamiltonian~(\ref{Hamilt}) is reduced to the
related single electron problem of the propagation of
the electron in a random interfacial potential
\begin{eqnarray}
\hat V & = & \sum_{n; \alpha=1,2}|\rho_n^{\alpha}\rangle \left(
\hat v_n^{(\rm el)\alpha} + \hat v^{(\rm sf) \alpha}_n
\right)\langle \rho_n^{\alpha}| 
\nonumber \\
& = & \sum_{n; \alpha=1,2}
|\rho_n^{\alpha}\rangle \hat v_n^{\alpha} \langle \rho_n^{\alpha}|,
\nonumber
\end{eqnarray}
where
$\hat v_n^{\alpha} = \hat v_n^{(\rm el)\alpha} + \hat v^{(\rm sf)\alpha}_n$ and 
the summation goes over the interface number $\alpha$ and over the sites $n$.
Here
\begin{equation}
\hat v_n^{(\rm el)\alpha} = \sum_{\mu=\uparrow,\downarrow} \gamma_n^{\alpha}
\left\{|s,\mu\rangle \langle d,\mu | + |d,\mu\rangle \langle s,\mu |  \right\}
\label{v_gamma}
\end{equation}
is the random potential of $s-d$ hybridization associated with
the site $n$ and
\begin{equation}
\hat v_n^{(\rm sf)\alpha} = J_n^{\alpha}\left\{
|s,\uparrow\rangle \hat S_{-}(\rho_n^{\alpha}) \langle s,\downarrow| +
|s,\downarrow\rangle \hat S_{+}(\rho_n^{\alpha}) \langle s,\uparrow|
\right\}
\label{v_spin}
\end{equation}
is the exchange-like interaction with the surface magnons.
The random quantities $\gamma_n^{\alpha}$ and $J_n^{\alpha}$ 
used here were introduced in Sec.~IIA.

  Now one can formulate the CPA by the ordinary way and the only difference
with respect to the usual situation of the bulk scattering 
is that the initial Green's functions have to be calculated for the trilayer system. 
We assume that the averaged propagator of the system
$G_{\mu}^{\alpha\beta}({\bf r}, {\bf r'})$ 
differs from the initial Green's function,
corresponding to the Hamiltonian $\hat H_0$ [see Eq.~(\ref{Hamilt})], 
by the self-energy correction
in Eqs.~(\ref{Green_eq}), (\ref{H_eff}). 
This means that the system behaves as if coherent potentials
$\Sigma_{\mu}^{\alpha\beta}$ had been assigned to each site of the interface $a$ and $b$. 
After the introducing of the effective medium $\hat\Sigma^{\alpha}$,
each site $\rho^{\alpha}_n$ becomes a source of the random 
potential $\hat u_n^{\alpha} = \hat v_n^{\alpha} - \hat\Sigma^{\alpha}$. 
The single-site $t$-matrix associated with potential $\hat u^{\alpha}_n$
is given by
\begin{equation}
\hat t_n^{\alpha} = \frac{1}
{1 - \{ \hat v_n^{\alpha} - \hat \Sigma^{\alpha} \} \hat G(z_{\alpha}) }
\{ \hat v_n^{\alpha} - \hat\Sigma^{\alpha} \},
\label{t_matrix}
\end{equation}
where $z_{\alpha}= a$ or $b$,
\begin{eqnarray}
\hat \Sigma^{\alpha} & = &  \sum_{\mu=\uparrow,\downarrow}\sum_{\beta,\gamma = s,d}
|\beta,\mu\rangle \Sigma_{\mu}^{\beta\gamma}(z_{\alpha}) \langle \gamma, \mu|,
\label{Self_E} \\
\hat G(z_{\alpha}) & = &  \sum_{\mu=\uparrow,\downarrow}\sum_{\beta,\gamma = s,d}
|\beta,\mu\rangle G_{\mu}^{\beta\gamma}(z_{\alpha}) \langle \gamma, \mu|. \nonumber
\end{eqnarray}
Here
\begin{equation}
G_{\mu}^{\beta\gamma}(z_{\alpha}) = \int_0^{\kappa_{max}}  G_{\mu}^{\alpha\beta}(z_{\alpha},\kappa)
\frac{\kappa d\kappa}{2\pi}
\label{G_kappa}
\end{equation}
is the averaged Green's function at the  interface ${\alpha}$ 
which is expressed via $\hat\Sigma^{\alpha}$ in accordance 
with Eq.~(\ref{G_int}).
The upper limit $\kappa_{max}$ is a cut-off of the in-plane momentum which
originates from the finite size of the Brillouin zone. 
For that, one has to substitute the Brillouin zone's
projection onto $(k_x, k_y)$-plane by a circle of 
the same square with a radius $\kappa_{max}$.
For the bcc lattice, $\kappa_{max} = 2\sqrt{\pi/\sqrt{2}}/a_0$, where $a_0$
is a lattice constant. 
The single-site $t$-matrix~(\ref{t_matrix}) is obviously different 
for different sites. At the same time it is supposed that scattering
by the random potential is taken into account in the averaged 
propagator $\hat G(z_{\alpha})$ by the self-energy operator $\hat \Sigma^{\alpha}$.
Therefore, we require that the ensemble average of the
single-site $t$-matrix vanishes, i.e.,
\begin{equation}
\langle \hat t_n^{\alpha}(\hat\Sigma^{\alpha}) \rangle  =
x{\langle \hat t_A^{\alpha}\rangle}_b  + y {\langle \hat t_B^{\alpha} \rangle}_b = 0.
\label{t_av}
\end{equation}
Here $t_A^{\alpha}$ and $t_B^{\beta}$ are the values of the single-site $t$-matrix
associated with a given site $n$ which is occupied by the A-type ion 
(ferromagnet's ion) or by the B-type ion (impurity), respectively.
Brackets ${\langle\dots\rangle}_b$ denote the averaging over magnon degrees of freedom.
The equation (\ref{t_av}) is the well-known self-consistent coherent potential
approximation (CPA) \cite{Soven,Ehrenreich} that implicitly determines the
self-energy operator $\hat\Sigma^{\alpha}$. The CPA equations (\ref{t_av}) 
are formulated for the particular case of electron scattering within 
the F/O interface. The straightforward calculations show that the single-site
$t$-matrix $\hat t^{\alpha}_n$ can be represented in the form:
\begin{equation}
\label{t_form}
\hat t_n^{\alpha} =
\left(
\begin{array}{cc}
\hat t_n^{\alpha\uparrow}(\hat n_{+}) & \hat t_n^{\alpha +}(\hat n_{-})\hat S_{-} \\
\hat t_n^{\alpha -}(\hat n_{+})\hat S_{+} & \hat t_n^{\alpha \downarrow}(\hat n_{-})
\end{array}
\right)
\end{equation}
with respect to spin-up and spin-down subspaces. Here 
$\hat n_{+} = \hat S_{+} \hat S_{-}$, 
$\hat n_{-} = \hat S_{-} \hat S_{+}$, and the blocks 
$\hat t_n^{\alpha\pm}$, $\hat t_n^{\alpha\uparrow(\downarrow)}$ 
are $(2 \times 2)$-matrices, functionally depending on 
$\hat n_{+}$ and $\hat n_{-}$, with the components designated
by indices of bands $s$ and $d$.

  To satisfy the condition~(\ref{t_av}), one has to consider
only the spin-conserving part of this equation as long as
$\langle t_n^{\alpha\pm} \hat S_{\mp}\rangle_b = 0$ since the expression
to be averaged contains unequal number of the creation and annihilation 
operators of the magnons. In order to calculate
$\langle t_n^{\alpha\uparrow(\downarrow)}\rangle_b$ 
we adopted the further approximation and assumed that 
\begin{eqnarray}
\langle t_n^{\alpha\uparrow}(\hat n_{+})\rangle_b 
& = & \hat t_n^{\alpha\uparrow}(n), 
\nonumber \\
\langle t_n^{\alpha\downarrow}(\hat n_{-})\rangle_b & = & \hat t_n^{\alpha\downarrow}(n),
\nonumber
\end{eqnarray}
here $n = n(T)$ is the average number of magnons
at the given temperature.
In other words, we substituted the operators
$\hat n_{\pm}$ by its averaged values. 
The function $n(T)$ is given by the
familiar expression
\begin{equation}
\label{n_T}
n(T) = \int\frac{d^2 {\bf q}}{(2\pi)^2}\frac{1}{e^{\beta\omega(\bf q)}-1}
\end{equation}
where $\omega({\bf q})$ is the spectrum of surface magnons. 
The approximation we made to take into account the spin-flip
processes is the static approximation for magnons.
The inelasticity of the electron-magnon processes may be taken 
into account in the energy conservation rule.
However, the characteristic magnon energy $\hbar\omega_0$ is much
less than the Fermi energy $\varepsilon_F$, and in the first
approximation one can neglect this energy $\hbar\omega_0$ ---
it becomes important in the case of finite voltages when the
process of emitting of a magnon by a hot electron
influences on the form of $I-V$ dependence. 
We, therefore, restricted our calculation to the case of 
small voltage bias with the voltage less than $\hbar\omega_0$.
Within this approximation, for the system of CPA equations 
we get
\begin{equation}
\begin{array}{ccc}
t_A^{\alpha\uparrow}(n)\,x + t_B^{\alpha\uparrow}(n)\,y &=& 0, \\
t_A^{\alpha\downarrow}(n)\, x + t_B^{\alpha\downarrow}\,(n) y &=& 0.
\end{array}
\label{CPA_system}
\end{equation}
As far as matrices $t_n^{\alpha\uparrow(\downarrow)}$ are $(2\times 2)$--blocks, 
thus for the general case Eq.~(\ref{CPA_system}) represents
the system of two matrix equations for 8 unknown quantities 
$\Sigma_{\mu}^{\beta\gamma}(z_{\alpha})$.
The system (\ref{CPA_system}) is one of the possible forms
of the CPA equations. But actually we used another representation 
of the CPA which was more convenient for the numerical 
implementation. For that we exploited the augmented-space
formalism by Mookerjee \cite{Mookerjee}, and the details
are presented in Appendix B.

\subsection{The vertex corrections}

\begin{figure}[t]
\begin{center}
\includegraphics[scale=0.3, angle=-90]{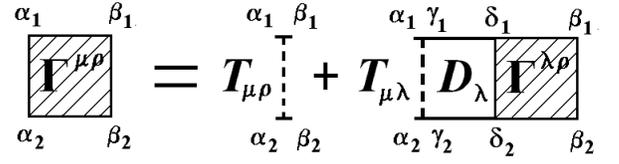}
\end{center}
\caption{The diagrammatic representation of the equation (see the text) 
corresponding to the calculation of the vertex part 
$\Gamma^{\mu\rho}$  in the "ladder" approximation.}
\end{figure}

  To find the vertex corrections we used the "ladder" approximation
combined with the CPA, that was originally proposed by
Velick\'y~\cite{Velicky}. The diagrammatic representation of 
the "ladder" approximation is given in Fig.\ 3. 
Since the scattering potentials on 
the different interfaces (at the points $a$ and $b$) 
do not correlate with each other, the vertex parts
$\Gamma^{\mu\rho}_{a(b)}
 \left({\scriptsize \begin{array}{cc}
 \beta_1 & \gamma_1\\
 \beta_2 & \gamma_2
 \end{array} }\right)
$ can be found independently for each interface and we will omit interface
suffix in the subsequent expressions. Let
$T^{\mu\rho}
 \left({\scriptsize \begin{array}{cc}
 \beta_1 & \gamma_1\\
 \beta_2 & \gamma_2
 \end{array} }\right) = 
\left \langle\left(t_n^{\alpha}\right)_{\mu\rho}^{*\beta_1\gamma_1}
\left(t_n^{\alpha}\right)_{\rho\mu}^{\gamma_2\beta_2} \right \rangle$ 
be the average of the product of two random $t$-matrices
over configurations and magnon distribution, where as before
$\mu,\rho$ are spin indices and $\beta,\gamma$ are orbital indices.
On the diagram in Fig.3, it is presented by the dashed line. The analytical
expression for the correlator
$
T^{\mu\rho}
 \left({\scriptsize \begin{array}{cc}
 \beta_1 & \gamma_1\\
 \beta_2 & \gamma_2
 \end{array} }\right)
$ can be found in accordance with the adopted approximate scheme of averaging
over the magnon degrees of freedom. Taking into account
Eq.~(\ref{t_form}) we obtain:
\begin{eqnarray}
T^{\uparrow\uparrow}
 \left({\scriptsize \begin{array}{cc}
 \beta_1 & \gamma_1\\
 \beta_2 & \gamma_2
 \end{array} }\right) & = &
x\,{t_A^{*\uparrow}(n)}^{\beta_1\gamma_1}{t_A^{\uparrow}(n)}^{\gamma_2\beta_2} 
\nonumber \\
& & + \ y\,{t_B^{*\uparrow}(n)}^{\beta_1\gamma_1}{t_B^{\uparrow}(n)}^{\gamma_2\beta_2},
\nonumber \\
T^{\downarrow\downarrow}
 \left({\scriptsize \begin{array}{cc}
 \beta_1 & \gamma_1\\
 \beta_2 & \gamma_2
 \end{array} }\right) & = &
x\,{t_A^{*\downarrow}(n)}^{\beta_1\gamma_1}{t_A^{\downarrow}(n)}^{\gamma_2\beta_2}
\nonumber\\
& & +\ y\,{t_B^{*\downarrow}(n)}^{\beta_1\gamma_1}{t_B^{\downarrow}(n)}^{\gamma_2\beta_2},
\\
T^{\uparrow\downarrow}
 \left({\scriptsize \begin{array}{cc}
 \beta_1 & \gamma_1\\
 \beta_2 & \gamma_2
 \end{array} }\right) & = &
x n\,{t_A^{*+}(n)}^{\beta_1\gamma_1}{t_A^{-}(n)}^{\gamma_2\beta_2}
\nonumber\\
& & +\ y n\,{t_B^{*+}(n)}^{\beta_1\gamma_1}{t_B^{-}(n)}^{\gamma_2\beta_2},
\nonumber \\
T^{\downarrow\uparrow}
 \left({\scriptsize \begin{array}{cc}
 \beta_1 & \gamma_1\\
 \beta_2 & \gamma_2
 \end{array} }\right) & = &
x n\,{t_A^{*-}(n)}^{\beta_1\gamma_1}{t_A^{+}(n)}^{\gamma_2\beta_2} 
\nonumber\\
& & + \ y n\,{t_B^{*-}(n)}^{\beta_1\gamma_1}{t_B^{+}(n)}^{\gamma_2\beta_2},
\nonumber
\end{eqnarray}
where $t^{\uparrow(\downarrow)}_{A(B)}(n)$ and
$t^{\pm}_{A(B)}(n)$ are the components of the single-site
$t$-matrix, Eq.~(\ref{t_form}). We also define the operator
\begin{eqnarray}
D^{\alpha}_{\mu}
 \left({\scriptsize \begin{array}{cc}
 \beta_1 & \gamma_1\\
 \beta_2 & \gamma_2
 \end{array} }\right) & = & \frac{1}{A}\sum_{\kappa}
 \left[G_{\kappa\mu}^{*\beta_1\gamma_1}(z_{\alpha})
  G_{\kappa\mu}^{\beta_2\gamma_2}(z_{\alpha})\right]
\nonumber\\
& & -\ G_{\mu}^{*\beta_1\gamma_1}(z_{\alpha})G_{\mu}^{\beta_2\gamma_2}(z_{\alpha})
\label{D}
\end{eqnarray}
denoting the propagator of the pair of electrons in the "ladder" equation
at the interface $z_{\alpha}$. Its definition follows from the fact that in
the diagram representation of the CPA the multiple scattering
on the given site is assumed to be incorporated into the single-site
$t$-matrix  $t_n^{\alpha}$, corresponding to the single vertex of any diagram.
Due to that, the subsequent sites in the "ladder" diagrammatic equation (Fig.3)
must not reproduce each other.  Therefore, the necessary correction is
subtracted in Eq.~(\ref{D}). The summation over $\kappa$ goes up to
$\kappa_{max}$ similar to Eq.~(\ref{G_kappa}). After that definitions the
analytical equation for the vertex part reads as
\begin{eqnarray}
\Gamma^{\mu\rho}
 \left({\scriptsize \begin{array}{cc}
 \alpha_1 & \beta_1\\
 \alpha_2 & \beta_2
 \end{array} }\right) & = &
T_{\mu\rho}
 \left({\scriptsize \begin{array}{cc}
 \alpha_1 & \beta_1\\
 \alpha_2 & \beta_2
 \end{array} }\right) + 
T_{\mu\lambda}
 \left({\scriptsize \begin{array}{cc}
 \alpha_1 & \gamma_1\\
 \alpha_2 & \gamma_2
 \end{array} }\right)
\label{Gamma_eq} \\
& & \times\ D_{\lambda}
 \left({\scriptsize \begin{array}{cc}
 \gamma_1 & \delta_1\\
 \gamma_2 & \delta_2
 \end{array} }\right)
\Gamma^{\lambda\rho}
 \left({\scriptsize \begin{array}{cc}
 \delta_1 & \beta_1\\
 \delta_2 & \beta_2
 \end{array} }\right). \nonumber
\end{eqnarray}
Eq.\ (\ref{Gamma_eq}) is the ordinary system of linear equations. 
It can be proved that the vertex parts found 
with the use of the "ladder" approximation 
and self-energies found with the use of the CPA
satisfy the Ward identity (\ref{Ward}), 
and hence the total tunneling current 
is a constant value $j(z) = j_0$, i.e.\ it does not
depend on $z$. The proof of the Ward identity is presented 
in Appendix~A.

\section{Results and discussion}

We considered the case of Fe/Al$_2$O$_3$/Fe tunnel junction.
The following parameters were chosen to describe the system.
According to estimations of Stearns \cite{Stearns}, for the itinerant
$s$--like electrons in Fe we set $k_{s}^{F\uparrow} = 1.09$~\AA$^{-1}$,
$k_{s}^{F\downarrow} = 0.42$~\AA$^{-1}$, and $m_s = 1.0\,m_e$
(here $m_e$ is bare electron mass). For the more localized $d$--electrons
(holes), we put $m_d = -10.0\,m_e$.
The $d$ density of states at the Fermi energy 
is larger for the minority spin band \cite{Moruzzi},
therefore, $k_{d}^{F\downarrow} > k_{d}^{F\uparrow}$, and we put
$k_{d}^{F\uparrow} = 0.45$~\AA$^{-1}$, $k_{d}^{F\downarrow} = 1.35$~\AA$^{-1}$.
The values of the Fermi momenta define the positions of band bottoms
$V^{\mu}_{1\alpha}$, $V^{\mu}_{3\alpha}$ ($\alpha = s,d$,
$\mu = \uparrow,\downarrow$) with respect to $\varepsilon_F$.
For $d$-electrons the Fermi momenta were chosen such a way that 
the interface density of states 
$\rho_{s(d)}^{\mu}(\varepsilon_F) = -1/{\pi}\,{\rm Im}G_{s(d)}^{\mu}(z_{\alpha})$
(where $z_{\alpha} = a,b$ are the positions of interfaces) 
comply with the following ratio 
$ \rho_s^{\downarrow(\uparrow)}(\varepsilon_F) : 
\rho_d^{\uparrow}(\varepsilon_F) :
\rho_d^{\downarrow}(\varepsilon_F) \sim 0.1 : 1 : 10$, which is the typical
situation for the case of $3d$ transition metals
(see, for a example, the calculations of Tsymbal and Pettifor \cite{Tsymbal}).
We also put the width of the barrier $w = 20$~\AA, and
$z_0 = a_0/2$ where $a_0 = 2.87$~\AA\ is a lattice
constant for bcc Fe ($z_0$ is a parameter describing the thickness
of the interface). 

  The main features of the band structure of $\alpha$-Al$_2$O$_3$ 
crystals were briefly presented in Sec.~IIA. In a view of that discussion, 
the following parameters of the model were taken to describe 
the amorphous Al$_2$O$_3$ barrier: the effective masses of electrons
and holes are $m_s^0 = 0.4\, m_e$, $m_h^0 = -10.0\, m_e$,
the positions of the conduction band bottom ($U_s$) and the top
of a valence band ($U_d$) are 
$U_s = - U_d = 3.0$~eV, i.e.\ the width of the band gap is $\approx 6.0$~eV,
and $\varepsilon_F$ is assumed to be a zero of energy.

In order to illustrate the general formalism presented in the
previous sections, let us first consider
the case of only $s-d$ impurity scattering (i.e.\ $T=0$ and there are no
spin-flip processes) when concentration of impurity ions 
on the interface is $(1-x)=0.5$. In this case all formulae
have a simple analytical form. 
As it was mentioned previously in Sec.~IIA, in the two-band
tight-binding model description of a binary alloy the random
variables are diagonal matrix elements of the Hamiltonian 
$\varepsilon_n^s(x)$, $\varepsilon_n^d(x)$ and
$\gamma_n(x)$ depending on whether the site $n$ is occupied by an 
A or B ion. To be more precise, one must consider the concentration 
dependence of these matrix elements. It reflects the fact that 
their values are modified by the existence of a charge transfer in the alloy.
Since in the framework of our model we are focusing on the 
hybridization effects (see Sec.~IIA for justification of this approach),
we have $\gamma_A(x) \ne \gamma_0^A$, $\gamma_B(x) \ne \gamma_0^B$ 
where $\gamma_0^A$ and $\gamma_0^B$ are hybridizations of the pure 
components. Let us assume for simplicity that in Eq.~(\ref{operators})  
$\gamma_0 = \bar\gamma = x\gamma_A + y\gamma_B = 0$.
Then for $x = 0.5$ we get $\gamma_A=-\gamma_B$,
and thus  $\gamma^2 = x(1-x)(\gamma_A - \gamma_B)^2$ is the
single parameter characterizing the amplitude of $s$-$d$ scattering
on the interface. The nonzero value
of $\bar\gamma$ will lead to a slight modification
of the $s$ and $d$-like eigenstates describing $s$ and $d$-like
electrons in the vicinity of the interface, 
and thus will not affect the qualitative results
presented below. We will omit spin suffixes since
we consider the spin-conserving scattering now.
At $x = 0.5$ only diagonal elements of the
self-energy matrix $\Sigma^{\alpha\beta}$ in the $sd$-space have
non-zero values. Then the Green's functions, e.g.\ at the point $a$, become
[see Eq.\ (\ref{G_int})\,]
\begin{eqnarray}
G^{ss}(a) & = & \int_0^{k_{max}} \biggl\{ 
 \frac{k_1^{\mu\,s}}{2m_s}(i - {\rm cotan}\,\varphi_1^s)
\nonumber \\
& & - \Sigma_{\mu}^{ss}(a) \biggr\}^{-1}\frac{\kappa d\kappa}{2\pi}, 
\label{G_a} \\
G^{dd}(a) & = & \int_0^{k_{max}} \biggl\{ 
\frac{k_1^{\mu\,d}}{2m_d}\left[i - {\rm cotan}\,(k_1^d z_0+\varphi_1^d)\right] 
\nonumber \\
& & -\ \Sigma_{\mu}^{dd}(a) \biggr\}^{-1}\frac{\kappa d\kappa}{2\pi}, 
\nonumber
\end{eqnarray}
and CPA equations are written as follows
\begin{equation}
\Sigma^{ss}=\frac{\gamma^2 G^{dd}}{1+\Sigma^{dd} G^{dd}}, \qquad
\Sigma^{dd}=\frac{\gamma^2 G^{ss}}{1+\Sigma^{ss} G^{ss}}
\label{CPA_50}
\end{equation}
which must be solved self-consistently
by means of converging iterative procedure.
As far as Green functions are diagonal, one can denote 
$\Gamma_{\alpha\beta} =
\Gamma \left({\scriptsize 
\begin{array}{cc}\alpha&\beta\\ \alpha& \beta\end{array} 
}\right)$
and can introduce the similar notations 
for quantities $T$ and $D$, defined in section IID.
Then one gets
\begin{equation}
T_{sd} = T_{ds} = \frac{\gamma^2}{\left|1+\Sigma^{dd} G^{dd}\right|^2} =
\frac{\gamma^2}{\left|1+\Sigma^{ss} G^{ss}\right|^2},
\label{T_sd}
\end{equation}
\[
T_{ss}=T_{dd}=0,
\]
\[
\Gamma_{ss} = \frac{T_{sd}^2 D^{dd}}{1 - T_{sd}^2 D^{ss} D^{dd}}, \quad
\Gamma_{dd} = \frac{T_{ds}^2 D^{ss}}{1 - T_{sd}^2 D^{ss} D^{dd}},
\]
\[
\Gamma_{sd} = \Gamma_{ds} = \frac{T_{sd}}{1 - T_{sd}^2 D^{ss} D^{dd}}.
\]
Since the mass of holes $m_h^0$ in the insulator 
is much larger than the electron mass $m_s^0$, 
the exponential factor $e^{-2q_2^d(b-a)}$ in formulae
(\ref{s_0_ab} -- \ref{s_vertex_b}) for tunneling conductance is negligibly
small as compared with one for $s$--like electrons. Therefore,
one may neglect the contribution from $d$--holes to the 
tunneling current. Then for the tunneling conductance we obtain
\[
\sigma =\sum_{\mu=\uparrow,\downarrow}\left( \sigma_b^{\mu} + \sigma_{\Gamma_a}^{\mu\mu}
+ \sigma_{\Gamma_b}^{\mu\mu}\right),
\]
where
\begin{equation}
\sigma_b^{\mu} = \frac{e^2}{2\pi\hbar}\int_{0}^{\kappa_{max}}
A^{ss}_{\kappa\mu}(a)\left(\frac{q_2^s}{m^0_s}\right)^2
A^{ss}_{\kappa\mu}(b)e^{-2 q_2^s w}\frac{\kappa d\kappa}{2\pi}
\label{s_ballistic}
\end{equation}
is the conductance corresponding to the "bubble" diagram 
(here $w = b - a$),
\begin{eqnarray}
\sigma_{\Gamma_a}^{\mu\mu}& = & 
\frac{e^2}{2\pi\hbar}\left(\int_0^{\kappa_{max}} A^{dd}_{\kappa\mu}(a)
\frac{\kappa d\kappa}{2\pi}\right)\,\Gamma^{(a)\mu\mu}_{\,\,ds}
\label{sa_diff} \\
&  \times &  \left[ \int_0^{\kappa_{max}} { \left| G_{\kappa'\mu}^{ss}(a) \right|}^2
{\left(\frac{q_2^s}{m^0_s}\right)}^2
A^{ss}_{\kappa'\mu}(b)e^{-2 q_2^s w}\frac{\kappa' d\kappa'}{2\pi}\right]
\nonumber \\
& + & \frac{e^2}{2\pi\hbar}\left(\int_0^{\kappa_{max}} A^{ss}_{\kappa\mu}(a)
\frac{\kappa d\kappa}{2\pi}\right)\,\Gamma^{(a)\mu\mu}_{\,\,ss}
\nonumber \\
& \times &  
\left[ \int_0^{\kappa_{max}} { \left| G_{\kappa'\mu}^{ss}(a) \right|}^2
{\left(\frac{q_2^s}{m^0_s}\right)}^2
A^{ss}_{\kappa'\mu}(b)e^{-2 q_2^s w}\frac{\kappa' d\kappa'}{2\pi}\right]
\nonumber
\end{eqnarray}
is the "vertex" contribution to the conductance on the left interface, and
\begin{eqnarray}
\sigma_{\Gamma_b}^{\mu\mu} & = &
\frac{e^2}{2\pi\hbar}\left(\int_0^{\kappa_{max}} A^{dd}_{\kappa\mu}(b)
\frac{\kappa d\kappa}{2\pi}\right)\,\Gamma^{(b)\mu\mu}_{\,\,ds}
\label{sb_diff} \\
& \times & \left[\int_0^{\kappa_{max}} { \left| G_{\kappa'\mu}^{ss}(b) \right|}^2
{\left(\frac{q_2^s}{m^0_s}\right)}^2
A^{ss}_{\kappa'\mu}(a)e^{-2 q_2^s w}\frac{\kappa' d\kappa'}{2\pi}\right]
\nonumber \\
& + & \frac{e^2}{2\pi\hbar}\left(\int_0^{\kappa_{max}} A^{ss}_{\kappa\mu}(b)
\frac{\kappa d\kappa}{2\pi}\right)\,\Gamma^{(b)\mu\mu}_{\,\,ss}
\nonumber \\
& \times & 
\left[\int_0^{\kappa_{max}} { \left| G_{\kappa'\mu}^{ss}(b) \right|}^2
{\left(\frac{q_2^s}{m^0_s}\right)}^2
A^{ss}_{\kappa'\mu}(a)e^{-2 q_2^s w}\frac{\kappa' d\kappa'}{2\pi}\right]
\nonumber
\end{eqnarray}
is the "vertex" contribution to the conductance
on the right interface. Here $A^{\alpha\alpha}_{\kappa\mu}$
are the transport densities of states which for the present case on
the left interface are given by the expressions
\begin{eqnarray}
A_{\kappa\mu}^{ss}(a) & = & \frac{k^{\mu s}_1/m_s}{{\left|
k_1^s(i - {\rm cotan}\, \varphi_1^s)/2m_s
- \Sigma^{ss}_{\mu}(a)\right|}^2},
\label{A_transport}  \\
A_{\kappa\mu}^{dd}(a) & = & \frac{k^{\mu d}_1/m_d}{{\left|
k_1^d[i - {\rm cotan}(k_1^d z_0 + \varphi_1^d)]/2m_d
- \Sigma^{dd}_{\mu}(a)\right|}^2}. \quad
\nonumber
\end{eqnarray}
and by the analogous expressions in the case of right interface.

We remind, that physically the "bubble" term in the total 
conductance is a contribution to the current due to
direct tunneling when the electron momentum $\kappa = k_{\|}$
in the plane of the interface is conserved. Scattering processes 
renormalize the "bubble" term with respect to the case of ballistic 
transport such that the effective "height" of the potential barrier seen by 
electron increases due to self-energy corrections $\Sigma^{ss}_{\mu}$
arising on the interfaces. The vertex contributions to the conductance
describe the tunneling assisted by interfacial scattering~--- 
that is the processes of tunneling with scattering on the F/O interface 
when the in-plane momentum is not conserved, i.e.\ $\kappa'\ne \kappa$
for the scattered electron. Thus, both momenta $\kappa$ and
$\kappa'$ in Eq.~(\ref{s_vertex}) determine the vertex corrections.

 Note also, that both vertex corrections $\sigma_{\Gamma_a}$
and $\sigma_{\Gamma_b}$ (\ref{sa_diff}), (\ref{sb_diff}) consist
of two terms with vertices $\Gamma_{sd}$ and $\Gamma_{ss}$.
The terms with vertex part $\Gamma_{ss}$ 
are contribution to the tunneling conductance 
from $s$-like electrons only. The scattering in the
$s$ channel described by $\Sigma^{ss}$ (\ref{CPA_50}) is caused 
initially by $s$-$d$ scattering. The terms with $\Gamma_{sd}$ 
describe either the process of diffuse scattering of $d$-like electron
to the $s$ state in the F/O interface and then tunneling of $s$-like
electron in the barrier, or the process of tunneling of $s$-electron
and then its scattering to $d$ state in the O/F interface and leaving
into the electrode.
 
 We have calculated the tunneling conductance and the TMR ratio
defined as ${\rm TMR} = ({\sigma_{P} - \sigma_{AP}})/{ \sigma_{AP} }$,
where $\sigma_P$ and $\sigma_{AP}$ are the total conductances
for the parallel (P) and antiparallel (AP) alignment of magnetic
moments in the F-layers. The CPA equations defining the self-energies
were solved numerically. The validity of the Ward identity, 
Eq.~(\ref{Ward}), was checked after the vertex parts $\Gamma$
had been computed at every step for a given value of the 
parameter $\gamma$. 

\begin{figure}[t]
\begin{center}
\includegraphics[scale=0.3, angle=0]{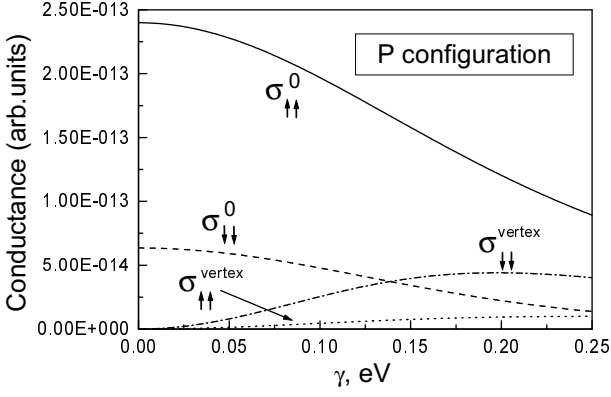}
\end{center}
\caption{"Bubble" and vertex contributions to the tunneling
conductance (in the units $e^2/2\pi\hbar$ per 
unit square 1\AA$^2$)
for the parallel (P) alignment of magnetic moments 
in the ferromagnetic layers
as a function of the scattering parameter $\gamma$
on the interface in the absence of spin-flip processes.
The parameters of the model are: $s$-like electrons~---
$k^{F\,s}_{\uparrow} = 1.09$\AA$^{-1}$,
$k^{F\,s}_{\downarrow} = 0.42$\AA$^{-1}$, $m_s = 1.0$, $m_s^0 = 0.4$;
$d$-like holes~---
$k^{F\,d}_{\uparrow} = 0.5$\AA$^{-1}$,
$k^{F\,d}_{\downarrow} = 1.4$\AA$^{-1}$, $m_d = m_h^0 = -10.0$,
the height of the potential barrier $U_s = - U_d = 3.0$~eV,
the width of the barrier $w = 20$~\AA,
the concentration of Fe ions on the interface $x=0.5$.
}
\end{figure}

The results are presented in Figs.\ 4--7.
In Fig.~4 the "bubble" (\ref{s_ballistic}) and the "vertex"
(\ref{sa_diff}), (\ref{sb_diff}) contributions to the conductance
for spin-up and spin-down channels are shown for the parallel (P) 
alignment of magnetic moments as a function of the scattering 
parameter $\gamma$ on the interface. (We remind that up to now only 
the spin-conserving scattering is considered.) As it was mentioned
previously, scattering suppresses the "bubble" conductance, 
and hence $\sigma_{\uparrow\uparrow}^0$ and $\sigma_{\downarrow\downarrow}^0$ 
are decreasing functions on $\gamma$.
The contribution from the "bubble" part
is larger for the majority spin $(\uparrow)$ channel since
$k^{Fs}_{\uparrow} > k^{Fs}_{\downarrow}$. On the contrary,
the contribution from the vertex corrections dominates for the 
minority spin $(\downarrow)$ channel --- that behavior can be 
understood as follows. 

 First, at small $\gamma^2$ the imaginary part of the self-energy 
$\Sigma^{ss}$, which describes the scattering of $s$-electrons,
behaves as ${\rm Im}\,\Sigma^{ss}_{\uparrow(\downarrow)} \sim
\pi\gamma^2 \rho^d_{\uparrow(\downarrow)}(\varepsilon_F)$
[see Eq.~(\ref{CPA_50})], where $\rho^d_{\uparrow(\downarrow)}(\varepsilon_F)$
are spin-up (down) $d$ density of states on the interface.
As far as $\rho^d_{\downarrow}(\varepsilon_F)$
is of the order of magnitude greater than $\rho^d_{\uparrow}(\varepsilon_F)$,
the scattering for $s$-like itinerant electrons from 
minority spin band ($\downarrow$) 
is more effective than for the majority spin ($\uparrow$) electrons.
The more effective scattering of spin-down $s$-electons
leads to the predominance of the minority spin contribution
with vertex part $\Gamma_{ss}^{\downarrow}$ over the majority spin
contribution with vertex $\Gamma_{ss}^{\uparrow}$, since at small
$\gamma$, $\Gamma_{ss}^{\uparrow(\downarrow)} \sim 
\gamma^4 D^{dd}_{\uparrow(\downarrow)}$ and 
$ D^{dd}_{\downarrow} \gg D^{dd}_{\uparrow}$, therefore 
$\Gamma_{ss}^{\downarrow} \gg \Gamma_{ss}^{\uparrow}$.
Second, the transport density of states  $A^{dd}(a)$ 
[Eq.\ (\ref{A_transport})] which determines the partial conductance
due to mixing of $s$ and $d$ channels in the interface
(terms with $\Gamma_{sd}$) is also larger for minority spin channel.
Thus, the combination of these factors results in domination 
of the vertex contribution to the conductance from the minority
spin channel. The results of calculations also showed that 
the $ss$ contribution to the vertex corrections 
$\sigma_{\Gamma}^{ss}$ (terms with vertices $\Gamma_{ss}$)
was much less than the $sd$ contribution $\sigma_{\Gamma}^{sd}$
(terms with $\Gamma_{sd}$).

\begin{figure}[t]
\begin{center}
\includegraphics[scale=0.3, angle=0]{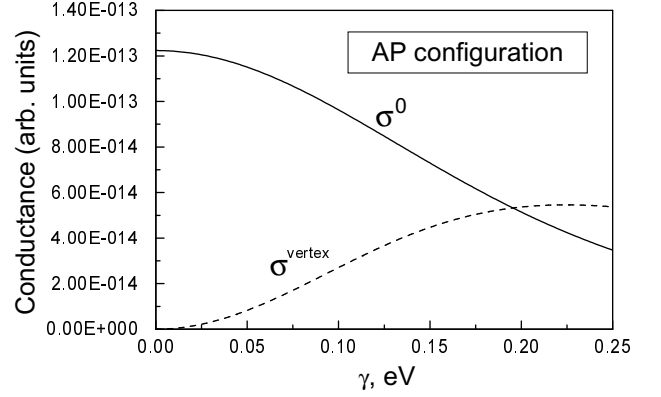}
\end{center}
\caption{"Bubble" and vertex contributions to the tunneling 
conductance (in the units $e^2/2\pi\hbar$ per 
unit square 1\AA$^2$)
for the antiparallel (AP) alignment of magnetic moments 
as a function of the scattering parameter $\gamma$ on the interface
in the absence of spin-flip processes. 
Parameters of the model are the same as in Fig.~4. }
\end{figure}

The "bubble" and vertex contributions to the conductance (which are the same for
spin-up and spin-down channels) for the antiparallel (AP) alignment of 
moments as a function of $\gamma$ are presented in Fig.~5. The total conductances for
P and AP configurations are shown in Fig.~6. 
Finally, in Fig.~7 the TMR ratio $ (\sigma_{P} - \sigma_{AP})/\sigma_{AP}$
as a function of scattering parameter $\gamma$ is shown.
When the amplitude of scattering is 
negligibly small ($\gamma = 0$), we have positive value of
the TMR~$\simeq 24\%$ --- that is the result of Slonczewski's theory \cite{Slonczewski}
under the chosen parameters for $s$--like electrons. With increasing of $\gamma$
the TMR amplitude is monotonically decreasing and can become even negative
if $\gamma > \gamma_c \approx 0.16$~eV.
 
To understand qualitatively the obtained results, conductance
and magnetoresistance can be rewritten in the general approximate form:
\begin{eqnarray}
\sigma_P\ & \approx & (\rho_{1}^{s\uparrow})^2 +  (\rho_{1}^{s\downarrow})^2 + 
\rho_{3}^{d\uparrow} \widetilde \Gamma^{\uparrow}_{sd} 
\rho_{1}^{s\uparrow} + 
\rho_{3}^{d\downarrow} \widetilde \Gamma^{\downarrow}_{sd} 
\rho_{1}^{s\downarrow},
\nonumber \\
\sigma_{AP} & \approx & 2 \rho_{1}^{s\uparrow} \rho_{1}^{s\downarrow}  + 
\rho_{3}^{d\uparrow} \widetilde \Gamma^{\uparrow}_{sd} 
\rho_{1}^{s\downarrow} + 
\rho_{3}^{d\downarrow} \widetilde \Gamma^{\downarrow}_{sd} 
\rho_{1}^{s\uparrow},
\nonumber \\
\Delta\sigma\ & = & \sigma_P - \sigma_{AP}\
\label{Dsigma} \\
& \approx &
(\rho_{1}^{s\uparrow} - \rho_{1}^{s\downarrow})^2 + 
(\rho_{1}^{s\uparrow} - \rho_{1}^{s\downarrow}) 
(\rho_{3}^{d\uparrow} \widetilde \Gamma^{\uparrow}_{sd} 
- \rho_{3}^{d\downarrow} \widetilde \Gamma^{\downarrow}_{sd}), 
\nonumber
\end{eqnarray}
where we took into account that $\sigma^{ss}_{\Gamma} \ll \sigma^{sd}_{\Gamma}$, 
and for high and thick enough barrier the main contribution 
to the tunneling conductance is due to electrons with momentum
almost perpendicular to the barrier (factors $e^{-2q^s_2w}$).
Consequently,  $\rho_{1}^{s\uparrow(\downarrow)}$ are the quasi-one-dimensional
$s$ density of states near the interface, and $\rho_{3}^{d\uparrow(\downarrow)}$
are the corresponding three-dimensional $d$ density of states. 
The $\widetilde\Gamma$ are renormalized vertex corrections.
In expression (\ref{Dsigma}) for $\Delta\sigma$, the first term due to 
direct tunneling is always positive. Concerning the second term, 
$\rho_{1}^{s\uparrow} > \rho_{1}^{s\downarrow}$ since 
$k_F^{s\,\uparrow} > k_F^{s\,\downarrow}$, but 
$\rho_{3}^{d\downarrow} \gg \rho_{3}^{d\uparrow}$ and, therefore, this
term is negative. This contribution decreases the magnetoresistance when
scattering parameter $\gamma$ becomes larger. At $\gamma > \gamma_c$
the $sd$ contribution overlaps the contribution
to $\Delta\sigma$ from the $ss$ channel, and we have 
${\rm TMR} < 0$. Thus, the inverse TMR ratio arisen in our model 
is caused by extremely strong scattering of negatively polarized 
$d$-like electrons (which give the indirect contribution to
the tunneling current) to the $s$-band on the interface.

\begin{figure}[t]
\begin{center}
\includegraphics[scale=0.29, angle=0]{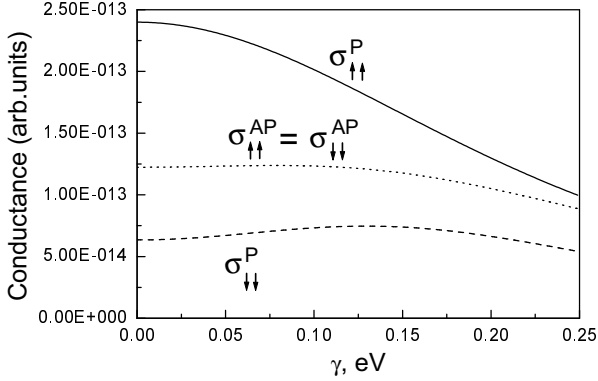}
\end{center}
\caption{The tunneling conductances of the individual spin channels
(in the units $e^2/2\pi\hbar$ per 
unit square 1\AA$^2$)
for the parallel (P) and antiparallel (AP) alignment of magnetic moments
as the function of the scattering parameter $\gamma$
on the interface in the absence of spin-flip processes.
The parameters are the same as in Fig.~4.
}
\end{figure}

 The parameter $\gamma$ determining the amplitude of $s$-$d$
scattering on the interface is defined as 
$\gamma^2 = x(1-x)(\gamma_A - \gamma_B)^2$.
One can regard the difference between hybridizations, $\gamma_A - \gamma_B$, 
as approximately a constant value for given constituents.
However, $(1-x)$ is a concentration
of impurity centers, and thus $\gamma$ is a measure of the imperfection
of the F/O interface. The proposed model, therefore, explains 
qualitatively the strong interface sensitivity of the tunneling
magnetoresistance effect \cite{Meservey, Mathon_JMMM, Moodera_Rev}.
According to Heine's discussion of the hybridization in transition 
metals \cite{Heine}, the hybridization constants are from about 1.0 
to 3.0 eV for different elements. One can assume that the difference 
$(\gamma_A - \gamma_B)$ is of the order of magnitude smaller
and, therefore, the critical value of scattering parameter 
$\gamma_c \simeq 0.16$~eV that we obtained is more or less realistic.
 
 The strong reduction of the TMR due to nonideal structure of the metal/insulator
interfaces is a well-known observation. The oxidation of a thin Al layer
leads to the undesirable oxidation of few metal monolayers close
to the F/O interface and thus to the formation of F-O oxides
(Fe$_3$O$_4$ \cite{Mitsuzuka}, CoO and Co$_3$O$_4$ \cite{JSMoodera}) --- 
that reduces the TMR \cite{Mitsuzuka,JSMoodera}.
On the other hand, if the too thick Al layer is not oxidized completely, 
the contamination of the interface by Al ions 
also reduces the TMR \cite{Mitsuzuka,JSMoodera}. 
The dependence of the TMR {\it vs.}\
the thickness of the Al overlayer has, therefore, a maximum and
the best TMR values achieved by Moodera's group lie
in the range $10~\div 16$~\AA\ \cite{JSMoodera}. 

\begin{figure}[t]
\begin{center}
\includegraphics[scale=0.3, angle=0]{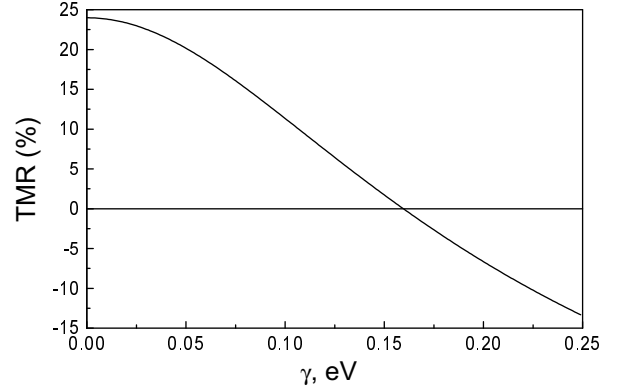}
\end{center}
\caption{The tunneling magnetoresistance (TMR) as a function of the 
scattering parameter $\gamma$ 
in the case of only elastic scattering on the interfaces.
The parameters are the same as in Fig.~4.
}
\end{figure}

  The contamination of the F/O interface by OH ions
in the early experiments by Merservey and Tedrow 
on tunneling with the superconductors \cite{Meservey}
led to the small measured values of a 
spin polarization $P$ for Ni and Gd.
The contamination was due to oxidizing of the Al films in the
laboratory air. The improved technique of samples preparation  
in a pure oxygen increased the values of $P$ for Ni and Gd, 
and for some rare-earth metals \cite{Meservey, Mathon_JMMM}.
 
 In a recent work by LeClair {\it et al.}\ \cite{LeClair_1}
the strong suppression of magnetoresistance was observed
in Co/Al$_2$O$_3$/Co tunnel junctions with a very thin Cr interfacial layer.
It was found that the TMR decayed exponentially on the 
Cr interlayer thickness 
with a length scale\ $\sim 1$\AA\ (approximately 0.5 monolayers). 
With the addition of only 3\AA\ Cr ($\approx 1.5$ monolayers)
the reduced TMR was only 10\% of the initial value for a control
junction. LeClair {\it et al.}\ \cite{LeClair_1}
presented some qualitative arguments that the suppression 
of a spin polarization (and, hence, the reduction of TMR) was 
due to more strongly suppression of majority $s$-$p$ density of states
compared with minority spins caused by the resonant scattering of majority
spin $sp$-electrons with the Cr $d$ states. From the point of view of our model, 
we can explain the strong degradation of the TMR by the strong electron 
scattering within the interfacial Co-Cr alloy that is formed 
under the preparation of extremely thin ($\sim 1$\ monolayer) 
Cr interlayer.

\begin{figure}[t]
\begin{center}
\includegraphics[scale=0.3, angle=0]{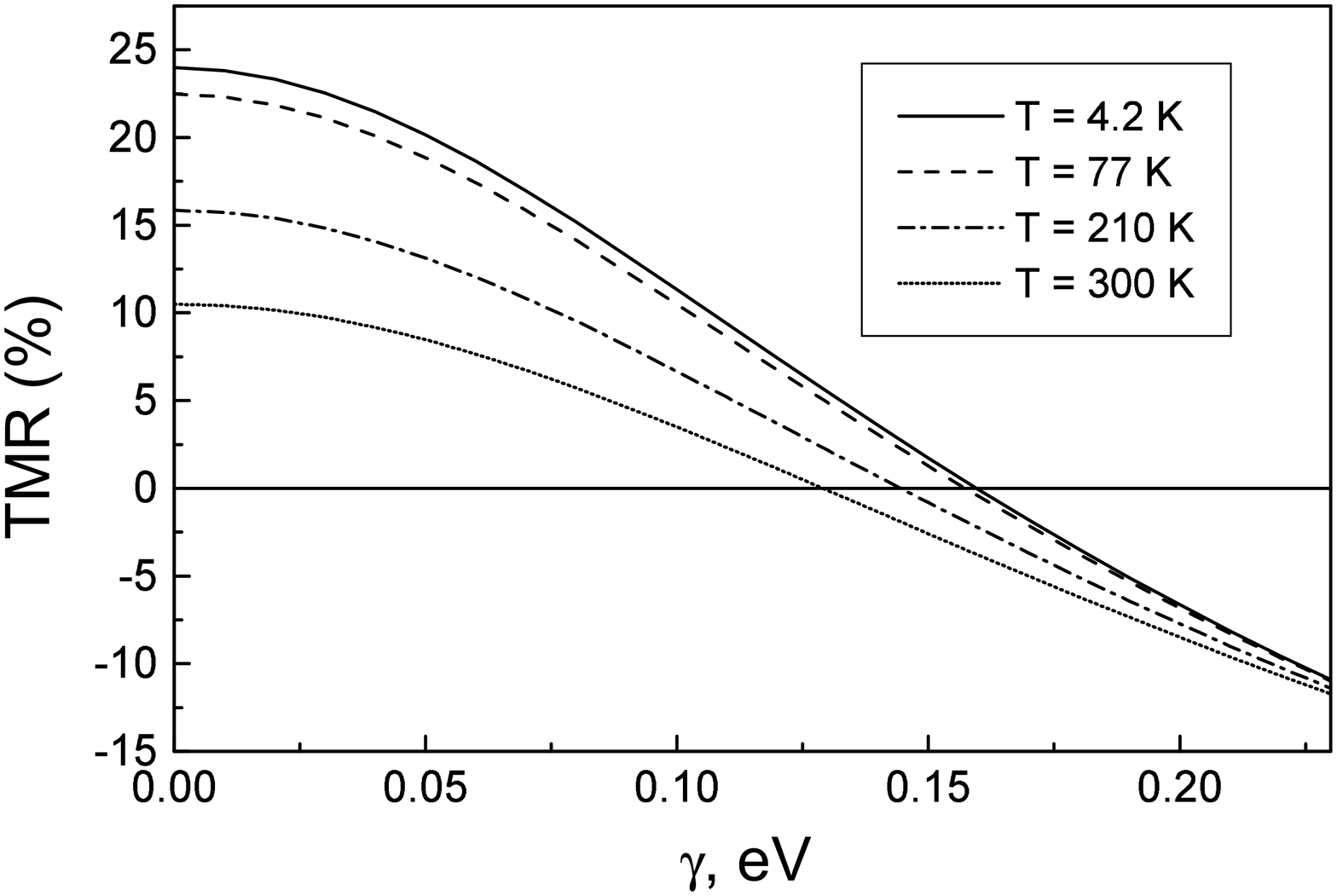}
\end{center}
\caption{The TMR for different temperatures as a function
of the scattering parameter $\gamma$. }
\end{figure}

\begin{figure}[b]
\begin{center}
\includegraphics[scale=0.35, angle=-90]{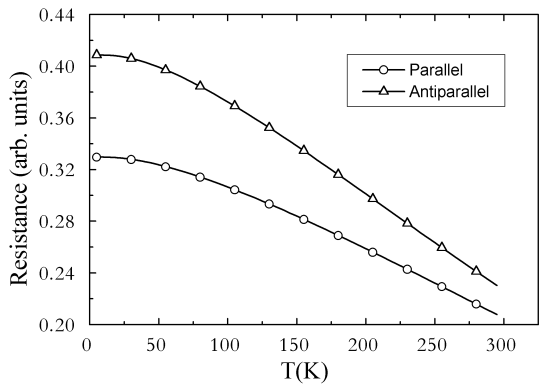}
\end{center}
\caption{The temperature dependence of the resistance for P and AP
configurations of the magnetic moments at $\gamma = 0$ and $x=50$\%.}
\end{figure}

    We have also calculated the temperature dependence of the TMR
taking into account spin-flip scattering in addition to
$s-d$ impurity scattering as it is described in details in sections IIC
and IID. For that, the average magnon number $n(T)$ (\ref{n_T})
as a function of the temperature was found in analogy with
Debye's treatment of phonons using the similar approach that was proposed
by S.~Zhang {\it et al} \cite{Levy}. The magnon dispersion 
relation in Eq.~(\ref{n_T}) was 
replaced by simple isotropic parabolic spectrum
\[
 \omega_{\bf q} = E_m\left(\frac{q}{k_{\rm max}}\right)^2,
\]
where $k_{\rm max}$ is the equivalent radius of the
two-di\-men\-si\-o\-nal Brillouin zone (see Eq.~\ref{G_kappa}),
and $E_m$ is related to Curie temperature $T_c$ and in
the mean-field approximation is given by
$E_m = 3k_B T_c/(S+1)$. For the chosen model of the dispersion
relation $w_{\bf q}$, one has to overcome the divergence on the lower limit
of the integral in Eq.~(\ref{n_T}). Therefore, one must introduce
a lower wave-length cutoff $E_c$ \cite{Levy}. Physically, it may represent
a finite coherence length due to interfacial roughness. In our
calculation we have taken the same parameters
that were used by S.~Zhang {\it et al.}\ \cite{Levy}\ for the analysis
of the zero-bias anomaly: $S = 3/2$, $k_B T_c = 110$~meV, and
$E_c = 4$~meV. Then, for the temperature range within the room
value, we have
\[
n(T) = -\frac{1}{2}\left(\frac{k_B T}{E_m}\right)\log
\left(1 - e^{-E_c/k_B T} \right).
\]
We have put $x=50\%$, $J_A = 2.0$~eV (for Fe atoms) and
$J_B = 0$~eV, i.e.\ it is supposed that spin-flip process
is possible, if an electron scatters on the Fe ion.
The TMR {\it vs.}\ the scattering parameter
$\gamma$ at different temperatures $T = 4.2, 77, 210$
and $300$~K are plotted in Fig.~8. In Fig.~9 the temperature
dependence of the resistance for the P and AP configuration
of magnetizations is presented for the same parameters, but $\gamma = 0$.
The results show that the TMR ratio decreases with increasing of
the temperature. Moreover, the resistances $R_{P}$ and $R_{AP}$
of the structure for both configurations, P and AP, are also decreasing
when temperature increases --- that is in the qualitative agreement 
with experimental data (e.g., see Fig.~4b in Ref.\ [38]). 
The physical mechanism of this effect
is related with the excitation of spin-flip processes in the system.
Due to these processes, the new channels of electron scattering appear
which are frozen at zero temperature. As the result, the conductance
of the system increases for both P and AP configurations and, therefore, 
the resistance drops.
The spin-flip processes mix the spin-up and spin-down
channels. Therefore, the relative difference
of the resistances decreases at different configurations
and the TMR decreases with increasing of temperature.

\section{Conclusions}

In conclusion, we would like to outline the main results obtained in 
the present work. Based on the analysis
of the band structure of 3$d$ ferromagnetic metals and Al$_2$O$_3$ crystals
and on the results of ab-initio calculations of the 
magnetoresistance for epitaxial tunnel junctions,
we built on an adaptation of the simplified two-band $s$-$d$ model 
to treat the diffuse electron transport in the non-ideal
F/O/F magnetic tunnel junctions. We had modeled the rough F/O 
interface by the random binary
alloy that is formed from the ions of the ferromagnet and impurities 
(e.g.\ the Al or O ions), and assumed that the main mechanism
of electron scattering on the interface which substantially
affects tunneling is the $s$-$d$ scattering. We used the Kubo formalism
to calculate the tunneling conductance and found the vertex
corrections to conductivity with the use of the "ladder" approximation 
combined with the CPA. The obtained results show that 
in the case of strong electron scattering within the interfacial alloy
the vertex corrections give the essential contribution to the 
tunneling conductance.  We proved that adopted approximations
lead to the physically correct results, namely, the non-local 
conductivity tensor is a constant function and, therefore, a tunneling
current is conserved. We showed that interfacial inter-band 
scattering substantially reduces a value of the TMR, which 
can become even negative in the case of extremely strong scattering.  
The reason of the suppression of the magnetoresistance
is the indirect contribution of negatively polarized $d$-like electrons 
to the tunneling current due to strong scattering to the $s$ band on 
the interface.  It is also shown that spin-flip electron scattering on the surface 
magnons within the interface leads to a further decrease of the TMR at 
finite temperature.  Thus, in the framework of the proposed 
model, we are able to explain qualitatively 
the strong interface sensitivity of the tunneling properties 
that is observed in the experiments.

\section*{Acknowledgments}

A.~Vedyayev and D.~Bagrets are grateful to CEA/\-Gre\-no\-ble/\-DRFMC/\-SP2M/\-NM
for fellowships, A.~Bag\-rets is grateful to 
Laboratoire Louis N{\'e}el/\-CNRS
(Grenoble, France) for hospitality. 
This work was partially supported by Russian Foundation
for Basic Research (grant No.~98-02-16806).

\appendix

\section{Ward identity}

In this Appendix we briefly describe how to obtain Eq.~(\ref{Ward}) (Ward identity),
and using a simple example we will show how this identity can be proved.

 Starting from Eq.~(\ref{current0}), one first has to compute the derivation
of the "bubble" conductivity. Assuming, that
the self-energy $\Sigma_{\mu}^{\alpha\beta}$
is symmetric with respect to the rearrangement of band indices, i.e.\
$\Sigma_{\mu}^{\alpha\beta} = \Sigma_{\mu}^{\beta\alpha}$, and 
using the fact that functions
$\psi_i^{\alpha}(z')$ are the solutions of the Schr{\"o}dinger equation
(\ref{Schrod}), for the derivation of the current matrix we obtain:
\[
\frac{\partial}{\partial z'}j^{\psi}_{\mu}(z') = \sum_{\alpha\beta = s,d}
2\, {\rm Im}\, \Sigma_{\mu}^{\alpha\beta}(z')[\rho_{\mu}^{\psi}(z')]^{\alpha\beta},
\]
where
\[
\Sigma_{\mu}^{\alpha\beta}(z') = \Sigma_{\mu}^{\alpha\beta}(a)\delta(z'-a) +
\Sigma_{\mu}^{\alpha\beta}(b)\delta(z'-b),
\]
and the matrix $[\rho_{\mu}^{\psi}(z')]^{\alpha\beta}$ is defined by Eq.~(\ref{rho}).
Then, from Eq.~(\ref{s_0}), (\ref{Lambda}) we have
\begin{eqnarray}
\displaystyle
\frac{\partial}{\partial z'}\sigma^0_{\mu}(z,z') & = &  
-\ \frac{e^2}{2\pi\hbar A}\sum_{\kappa}\sum_{\alpha\beta = s,d} \Bigl[
2\,{\rm Im}\, \Sigma_{\mu}^{\alpha\beta}(a)
\nonumber \\
& &\times\ \delta(z'-a) \Lambda^{\alpha\beta}_{\mu\kappa}(z,a)
\label{App_s0} \\
& & + \ 2\, {\rm Im}\, \Sigma_{\mu}^{\alpha\beta}(b)
\delta(z'-b) \Lambda^{\alpha\beta}_{\mu\kappa}(z,b)\Bigr]. 
\nonumber
\end{eqnarray}
To proceed further, let us compute the derivation of the vertex correction to
conductivity. The initial expression for the matrix $\Lambda_{\mu\kappa}$
(see diagram in Fig.~2) is written as
\[
\Lambda^{\gamma_1\gamma_2}_{\mu\kappa}(a,z') = \sum_{\alpha=s,d}
G^{*\gamma_1\alpha}_{\mu\kappa}(a,z')
\frac{\stackrel{\leftrightarrow}{\nabla}_{z'}}{2im_{\alpha}}
G^{\alpha\gamma_2}_{\mu\kappa}(z',a).
\]
Taking into account that the Green's functions are the solution of the
Eq.~(\ref{Green_eq}), we obtain
\begin{eqnarray}
\frac{\partial}{\partial z'}\Lambda^{\gamma_1\gamma_2}_{\mu\kappa}(a,z') & = &
 - 2 \delta(z'-a)\, {\rm Im}\, G_{\mu\kappa}^{\gamma_1\gamma_2}(a)
\nonumber
\\
& & +\ 2 \sum_{\alpha\beta = s,d}\,{\rm Im}\Sigma_{\mu}^{\alpha\beta}(a)
G^{*\alpha\gamma_1}_{\mu\kappa}(a) 
\nonumber \\
& & \times \ G^{\beta\gamma_2}_{\mu\kappa}(a) \delta(z'-a).
\nonumber
\end{eqnarray}
Substituting the obtained expression to formula (\ref{s_vertex})
for the vertex correction, we get
\begin{eqnarray}
\displaystyle
\frac{\partial}{\partial z'}\sigma^{\Gamma_a}_{\mu\rho}(z,z') & = &
\frac{e^2}{2\pi\hbar A^2}\sum_{\kappa\kappa'}2\delta(z'-a)
\Lambda_{\mu\kappa}^{\beta_1\beta_2}(z,a)
\nonumber \\
& & \times\ \Gamma^{\mu\rho}
\left({\scriptsize \begin{array}{cc}
 \beta_1 & \gamma_1\\
 \beta_2 & \gamma_2
 \end{array} }\right)
\Bigl[{\rm Im}\,G_{\rho\kappa'}^{\gamma_1\gamma_2}(a) 
\label{App_s_vertex} \\
& & -\ {\rm Im}\,\Sigma_{\rho}^{\alpha_1\alpha_2}(a)
G^{*\alpha_1\gamma_1}_{\rho\kappa'}(a) G^{\alpha_2\gamma_2}_{\rho\kappa'}(a) \Bigr].
\nonumber
\end{eqnarray}
The summation here is also performed 
over indices $\alpha_i$, $\beta_i$, and $\gamma_i$.
A similar expression can be written for the derivation of the vertex correction
$\sigma^{\Gamma_b}_{\mu\rho}(z,z')$ at interface $z=b$.
From (\ref{App_s0}) and (\ref{App_s_vertex})
one can obtain the final expression (\ref{Ward})
for Ward identity.

   Let's prove identity for the simple case of only
$s-d$ scattering, when there are no spin-flip processes,
and $x = 0.5$. This situation was considered in Sec.~III,
and there were introduced the notations $\Gamma_{\alpha\beta}$,
$T_{\alpha\beta}$, and $D^{\alpha\beta}$ ($\alpha\beta = s,d$)
for the components
of the $(2\times 2)$--matricies $\Gamma$, $T$, and $D$. Within 
these notations, the identity (\ref{Ward}), which has to be proved, 
can be written as
\[
{\rm Im}\,\Sigma^{\alpha\alpha} = \sum_{\beta=s,d}\Gamma_{\alpha\beta}
\left[
{\rm Im}\,G^{\beta\beta} - \frac{1}{S}\sum_{\kappa}\left|G_{\kappa}^{\beta\beta}\right|^2
{\rm Im}\,\Sigma^{\beta\beta}
\right].
\]
Here we omitted spin suffixes, 
all values associated either with the interface $a$ or $b$, 
$G^{\beta\beta}$ and $\Gamma_{\alpha\beta}$ are defined by
Exps.~(\ref{G_a},\ \ref{CPA_50}).
From (\ref{D},\ \ref{Gamma_eq}) it follows that
\[
D^{\alpha\alpha} = \frac{1}{A}\sum_{\kappa}
\left|G_{\kappa}^{\alpha\alpha}\right|^2 -
\left|G^{\alpha\alpha}\right|^2, \quad
\Gamma^{-1} = T^{-1} - D.
\]
Using these expressions, Ward identity can be written in the form:
\[
{\rm Im}\,G^{\alpha\alpha} -
\sum_{\beta = s,d}
\Bigl\{[T^{-1}]_{\alpha\beta} + \delta_{\alpha\beta}|G^{\alpha\alpha}|^2
\Bigr\}
{\rm Im}\,\Sigma^{\beta\beta} = 0.
\]
Using (\ref{T_sd}), we then can find that 
\begin{eqnarray}
{\rm Im}\,\Sigma^{ss} & = & \frac{1}{\rm Den}
\Bigl(
|G^{dd}|^2{\rm Im}\,G^{ss} 
\nonumber \\
& & +\ \gamma^{-2}|1 + \Sigma^{ss}G^{ss}|^2 {\rm Im}\, G^{dd}
\Bigr),
\nonumber \\
{\rm Im}\,\Sigma^{dd} & = & \frac{1}{\rm Den}
\Bigl(
|G^{ss}|^2{\rm Im}\,G^{dd} 
\nonumber \\
& & +\ \gamma^{-2}|1 + \Sigma^{dd}G^{dd}|^2 {\rm Im}\, G^{ss}
\Bigr),
\nonumber
\end{eqnarray}
where
\[
{\rm Den} = |G^{ss}|^2|G^{dd}|^2 -
\gamma^{-4}|1 + \Sigma^{ss}G^{ss}|^4.
\]
The same expessions for ${\rm Im}\,\Sigma^{ss}$ and
${\rm Im}\,\Sigma^{dd}$ follow directly from the CPA equations (\ref{CPA_50}) ---
that finishes the proof.

\section{The derivation of the CPA equations}

In this Appendix we will derive the CPA equations for our particular case
using the augmented-space formalism (ASF) \cite{Mookerjee}.
As was described previously, we assume the F/O interface to be a random
binary alloy of the type A$_x$B$_{1-x}$, where A are ions of the
ferromagnet and B are impurities. 
Following the ASF, we associate each random variable $\gamma_n^{\alpha}$ and $J_n^{\alpha}$
with the self-conjugate operators $\tilde\gamma$ 
and $\tilde J$, respectively, which are determined in
the auxiliary 2-dimensional vector space $\Phi$ such a way, that the spectrum of these operators
coincides with the set of possible values of random variables. For the sake of clarity, hereafter,
the sign "tilda"  is ascribed to any operator acting on the auxiliary space.
We also define the
orthonormalized basis $|s\rangle$, where $s=A$ or $B$, which are eigenvectors 
of $\tilde\gamma$ and $\tilde J$, so that
\begin{eqnarray}
\tilde\gamma|A\rangle = \gamma_A|A\rangle,& \qquad \tilde J|A\rangle = J_A|A\rangle, \nonumber \\
\tilde\gamma|B\rangle = \gamma_B|B\rangle,& \qquad \tilde J|B\rangle = J_B|B\rangle.  \label{Phi}
\end{eqnarray}
According to that definition, $\tilde\gamma$ and $\tilde J$
commutate with each other.
Let now $f(\gamma_n^{\alpha},J_n^{\alpha})$
be a function or an operator of random variables $\gamma_n^{\alpha}$ and $J_n^{\alpha}$. 
Then, the operator in the auxiliary space $\Phi$, associated with function $f$, 
is defined as $\tilde f = f(\tilde \gamma, \tilde J)$ and according to~(\ref{Phi}), e.g.
$\langle A|\tilde f|A \rangle$ is a value of $f$, if the site $n$ is occupied by an ion A.
One can introduce another orthonormal basis in $\Phi$
\[
   |0\rangle = \sqrt{x}|A\rangle + \sqrt{y}|B\rangle, \qquad
   |1\rangle = \sqrt{y}|A\rangle - \sqrt{x}|B\rangle,
\]
so that the operators $\tilde\gamma$ and $\tilde J$ in this representation are written as
\begin{equation}
  \tilde \gamma = \left(
\begin{array}{cc}
  \gamma_0 & \gamma \\
  \gamma & \gamma_1
\end{array} \right), \qquad
 \tilde J = \left(
\begin{array}{cc}
  J_0 & \delta \\
  \delta & J_1
\end{array} \right).
\label{operators}
\end{equation}
Here $x$ and $y$ are the concentration of A-type ions 
(ferromagnet's ions) and B-type ions (impurities) on the interface,
respectively, and
\begin{eqnarray}
\gamma_0 = x\,\gamma_A + y\,\gamma_B,& \quad & \gamma_1 = y\,\gamma_A + x\,\gamma_B,
\nonumber \\
\gamma = \sqrt{xy}(\gamma_A - \gamma_B),& \quad &  J_0 = x J_A + y J_B,
\nonumber \\
\quad J_1 = y J_A + x J_B,& \quad & \delta = \sqrt{xy}(J_A - J_B).
\nonumber
\end{eqnarray}
Then one can prove that the average value of $f$ is given by $\bar f = \langle 0|\tilde f|0\rangle$.
Together with Eq.~(\ref{operators}) this property is the way to evaluate the average
of any given operator, depending on the random variables, and 
we can apply this method to average the $t$-matrix~(\ref{t_matrix}).

Following the general scheme of the ASF, 
outlined above, the random effective potential
$\hat u_n^{\alpha} = \hat v_n^{\alpha} - \hat\Sigma^{\alpha}$ is associated with the
operator $\tilde U-\tilde\Sigma$ acting in the augmented vector space  $\Phi\otimes L$, where $L$ denotes
the 4-dimensional space of orbital ($s,d$) and spin ($\uparrow,\downarrow$) electron
degrees of freedom. In accordance with~(\ref{v_gamma}), (\ref{v_spin}) and (\ref{operators})
this operator has the form
\begin{eqnarray}
\tilde U -\tilde\Sigma &=& \left(
\begin{array}{cc}
\hat U_0-\hat\Sigma & \hat \Delta \\
\hat \Delta & \hat U_1-\hat\Sigma
\end{array}
\right) \nonumber \\
 &=& \left(
\begin{array}{cc}
\hat\gamma_0 +\hat J_0 \hat v - \hat\Sigma & \hat\gamma + \hat\delta\hat v \\
\hat\gamma + \hat\delta\hat v & \hat\gamma_1 + \hat J_1 \hat v - \hat\Sigma
\end{array}
\right)
\label{U}
\end{eqnarray}
where $\hat\Sigma$ is defined by~(\ref{Self_E}), operators $\hat\gamma_{(i)}$ are defined
similar to~(\ref{v_gamma}), and other operators are given by
\begin{eqnarray}
\hat v & = &
|\uparrow\rangle \hat S_{-}(\rho_n^{\alpha}) \langle\downarrow| +
|\downarrow\rangle \hat S_{+}(\rho_n^{\alpha}) \langle\uparrow|, \nonumber \\
\hat \delta &= & \delta |s\rangle\langle s|, \quad
\hat J_i = J_i|s\rangle\langle s| \quad (i = 0,1).
\nonumber
\end{eqnarray}

Let us also introduce the nonrandom averaged propagator acting in 
the augmented space
\[
\tilde G = \left(
\begin{array}{cc}
\hat G & 0 \\
0 & \hat G
\end{array}
\right),
\]
and associated with potential $\tilde U$ (\ref{U}) the augmented scattering
$t$-matrix
\begin{equation}
\tilde t = \left(\tilde U -\tilde\Sigma\right)\frac{1}
{1-\tilde G \left(\tilde U-\tilde\Sigma\right)} = \left(
\begin{array}{cc}
\hat t_{00} & \hat t_{01} \\
\hat t_{10} & \hat t_{11}
\end{array}
\right).
\end{equation}
Its projection onto the zero-level $|0\rangle$ of the augmented space
$\hat t_{00} = \langle 0|\tilde t|0\rangle$ coincides with the average from
the "physical" random $t$-matrix~(\ref{t_matrix}).
The subsequent averaging over magnon degrees of freedom ${\langle t_{00} \rangle}_b$
must vanish due to condition (\ref{CPA_system}).

  To proceed further, let's introduce the electron propagator
$\hat G_1$
\[
\hat G_1 = \frac{1}{\hat G^{-1} - ( \hat U_1 - \hat\Sigma ) } =
\left(
\begin{array}{cc}
\hat G_1^{\uparrow} & \hat G_1^{+} \\
\hat G_1^{-} & \hat G_1^{\downarrow}
\end{array}
\right),
\]
which is associated with the "propagation" of an electron on the first level of
the augmented space in the potential $\hat U_1-\hat\Sigma$.
Taking into account the explicit form of $\hat U_1 - \hat\Sigma$ with respect to 
spin-up and spin-down subspaces,
\[
\hat U_1-\hat\Sigma =\left(
\begin{array}{cc}
\hat\gamma_1 - \hat\Sigma^{\uparrow} & \hat J_1\hat S_{-} \\
\hat J_1\hat S_{+} & \hat\gamma_1 - \hat\Sigma^{\downarrow}
\end{array}
\right)
\]
where operators $\hat\gamma_1$ and $\hat J_1$ are 
\begin{eqnarray}
\hat\gamma_1 & = & \gamma_1 \{ |s\rangle \langle d| + |d \rangle \langle s| \},
\nonumber \\
\hat J_1 & = & J_1 |s\rangle \langle s|, \qquad \qquad 
\end{eqnarray}
one gets
\begin{eqnarray}
\hat G_1^{\uparrow} &=& \left[ 1 - \hat G^{\uparrow}
\left(\hat\gamma_1 - \Sigma^{\uparrow} +
\hat J_1\hat n_{+}\hat g_1^{\downarrow}\hat J_1 \right)
\right]^{-1}\hat G^{\uparrow},  \nonumber \\
\label{G_1}
\hat G_1^{\downarrow} &=& \left[ 1 - \hat G^{\downarrow}
\left(\hat\gamma_1 - \Sigma^{\downarrow} +
\hat J_1\hat n_{-}\hat g_1^{\uparrow}\hat J_1 \right)
\right]^{-1}\hat G^{\downarrow}, \\
\hat G_1^{+} &=& \hat G_1^{\uparrow}\hat J_1 \hat S_{-} \hat g_1^{\downarrow}, \quad
\hat G_1^{-} = \hat G_1^{\downarrow}\hat J_1 \hat S_{+} \hat g_1^{\uparrow}, \nonumber
\end{eqnarray}
where
\[
\hat g_1^{\uparrow(\downarrow)} = \left[ 1 - \hat G^{\uparrow(\downarrow)}
\left(\hat\gamma_1 - \hat \Sigma^{\uparrow(\downarrow)}\right)\right]^{-1}
\hat G^{\uparrow(\downarrow)},
\]
and $\hat n_{+} = \hat S_{-} \hat S_{+}$,
$\hat n_{-} = \hat S_{+} \hat S_{-}$. 
The physical meaning of these formulae is rather transparent.  The Green's
function $g_1^{\uparrow(\downarrow)}$ corresponds to 
the propagation of the electron in 
a spin conserving part of the potential $\hat U_1-\hat\Sigma$ which is
$\hat\gamma_1 - \hat \Sigma^{\uparrow(\downarrow)}$,
while $\hat G_1^{\uparrow{(\downarrow)}}$ corresponds to scattering on the potential
$\hat\gamma_1 - \hat \Sigma^{\uparrow(\downarrow)}
+ \hat J_1 \hat n_{\pm}\hat g_1^{\downarrow(\uparrow)} \hat J_1$, renormalized
with respect to the spin-conserving potential
due to the interaction with surface magnons on the interface.

   Coming back to the evaluation of scattering matrix element
$\hat t_{00}$, let introduce the "denominator" $\tilde D$, corresponding to 
the whole augmented potential $\tilde U-\tilde\Sigma$
\[
\tilde D = \frac{1}{1 - \tilde G \left(\tilde U - \tilde \Sigma\right) } =
\left(
\begin{array}{cc}
\hat D_{00} & \hat D_{01} \\
\hat D_{10} & \hat D_{11}
\end{array}
\right)
\]
Again, using the technique of the inversion of a matrix in the block form
and taking into account the elements $\tilde U$~(\ref{U}) with respect to
the auxiliary space $\Phi$, the blocks of $\tilde D$ can be expressed in terms
of the propagator $\hat G_1$ as follows
\begin{eqnarray}
\label{Den}
\hat D_{00} &=& \left[1-\hat G\left(\hat U_0 - \hat\Sigma +
\hat\Delta \hat G_1 \hat\Delta \right) \right]^{-1},  \nonumber \\
\hat D_{10} &=& \hat G_1 \hat\Delta \hat D_{00}, \quad
\hat D_{01} = \hat G_0 \hat\Delta \hat D_1,  \\
\hat D_{11} &=& \left(1 + \hat G_0 \hat\Delta \hat G_1 \hat\Delta  \right)\hat D_{00}, \nonumber
\end{eqnarray}
where we define the propagator $\hat G_0$,
\[
\hat G_0 = \left[1-\hat G\left(\hat U_0 - \hat\Sigma +
\hat\Delta \hat G_1 \hat\Delta \right) \right]^{-1} \hat G =
\left(
\begin{array}{cc}
\hat G_0^{\uparrow} & \hat G_0^{+} \\
\hat G_0^{-} & \hat G_0^{\downarrow}
\end{array}
\right),
\]
corresponding to the propagation of the electron in the effective potential
$\hat W-\hat\Sigma$, where
\begin{equation}
\label{W_eff}
\hat W = \hat U_0 + \hat\Delta\hat G_1 \hat\Delta =
\left(
\begin{array}{cc}
\hat w^{\uparrow} & \hat J_{+}\hat S_{-} \\
\hat J_{-}\hat S_{+} & \hat w^{\downarrow}
\end{array}
\right).
\end{equation}
The potential $\hat W$ can be regarded as a renormalization of the "virtual"
crystal potential $\hat U_0$ of the zero level of the augmented space, 
representing the average of the random potential on a site. The renormalization comes
from the "interaction" $\hat\Delta$  in the auxiliary space with the first
level, being described by the propagator $\hat G_1$. Using the explicit
form of $\hat\Delta$,
\[
\hat\Delta =\left(
\begin{array}{cc}
\hat\gamma & \hat\delta\hat S_{-} \\
\hat\delta\hat S_{+} & \hat\gamma
\end{array}
\right),
\]
one can write down the elements $\hat w^{\uparrow(\downarrow)}$
and $\hat J_{\pm}$:
\begin{eqnarray}
\hat w^{\uparrow(\downarrow)} & = & \hat\gamma_0 +
\hat\gamma \hat G_1^{\uparrow(\downarrow)}\hat\gamma  +
\hat\gamma \hat G_1^{\uparrow(\downarrow)}\hat J_1
\hat n_{\pm}\hat g_1^{\downarrow(\uparrow)}\hat\delta
\nonumber \\
& & +\ \hat\delta\hat G_1^{\downarrow(\uparrow)}
\hat J_1\hat n_{\pm}\hat g_1^{\uparrow(\downarrow)}\hat\gamma +
\hat\delta\hat n_{\pm}\hat G_1^{\downarrow(\uparrow)}\hat\delta,
\label{w_up_down}
\end{eqnarray}
\begin{eqnarray}
\hat J_{\pm} & = & \hat J_0 +
\hat\gamma \hat G_1^{\uparrow(\downarrow)}\hat\delta +
\hat\gamma \hat G_1^{\uparrow(\downarrow)}\hat J_1
\hat g_1^{\downarrow(\uparrow)}\hat\gamma
\nonumber \\
& & +\ \hat\delta\hat G_1^{\downarrow(\uparrow)}
\hat J_1\hat n_{\pm}\hat g_1^{\uparrow(\downarrow)}\hat\delta  +
\hat\delta\hat G_1^{\downarrow(\uparrow)}\hat\gamma.
\label{J_pm}
\end{eqnarray}
One can regard $\hat w^{\uparrow(\downarrow)}$ as the effective 
spin-conserving potential, taking into account the effects of a disorder, 
and $\hat J_{\pm}$ --- as the renormalized electron-magnon interaction.
Finally, the simple algebra using~(\ref{U}) (\ref{Den}), yields
\begin{eqnarray}
\hat t_{00} &=& \langle 0|(\tilde U - \tilde\Sigma)\tilde D|0\rangle
\nonumber \\
& = &  \left(\hat W - \hat \Sigma\right)\left[1-
\hat G \left(\hat W - \hat \Sigma \right)\right]^{-1} \nonumber \\
&=& \left(
\begin{array}{cc}
\hat t_0^{\uparrow} & \hat t_0^{+}\hat S_{-} \\
\hat t_0^{-}\hat S_{+} & \hat t_0^{\downarrow}
\end{array}
\right). \nonumber
\end{eqnarray}
Thus, we have obtained the logical result, that the $t$-matrix, 
averaged over configurations, corresponds to scattering on 
the effective potential $\hat W-\hat\Sigma$.

   To evaluate the spin-conserving $\hat t_0^{\uparrow(\downarrow)}$
and spin-flip $\hat t_0^{\pm}$ parts of scattering matrix, we introduce the
propagators
\begin{equation}
\hat g_0^{\uparrow(\downarrow)} = \left[ 1 - \hat G^{\uparrow(\downarrow)}
\left(\hat w^{\uparrow(\downarrow)} -
\hat \Sigma^{\uparrow(\downarrow)}\right)\right]^{-1}\hat G^{\uparrow(\downarrow)},
\label{g0_up_down}
\end{equation}
corresponding to the potentials
$\hat w^{\uparrow\downarrow} -\Sigma^{\uparrow\downarrow}$ and define the
"denominator"
\[
\hat D_0 = \left[1-\hat G\left(\hat W - \hat\Sigma\right)\right]^{-1}
=\left(
\begin{array}{cc}
\hat D_0^{\uparrow} & \hat D_0^{+} \\
\hat D_0^{-} & \hat D_0^{\downarrow}
\end{array}
\right).
\]
related with the effective potential $\hat W -\hat\Sigma$.
The elements of $\hat D_0$ and propagator $\hat G_0$ can be expressed
via $\hat g_0^{\uparrow(\downarrow)}$ using the form of
potential $\hat W$~(\ref{W_eff}), namely
\begin{eqnarray}
\hat G_0^{\uparrow} &=& \left[ 1 - \hat G^{\uparrow}
\left(\hat w^{\uparrow} - \Sigma^{\uparrow} +
\hat J_{+}\hat n_{+}\hat g_0^{\downarrow}\hat J_{-}\right)
\right]^{-1}\hat G^{\uparrow},  \nonumber \\
\nonumber
\hat G_0^{\downarrow} &=& \left[ 1 - \hat G^{\downarrow}
\left(\hat w^{\downarrow} - \Sigma^{\downarrow} +
\hat J_{-}\hat n_{-}\hat g_0^{\uparrow}\hat J_{+}\right)
\right]^{-1}\hat G^{\downarrow}, \\
\hat G_0^{+} &=& \hat G_0^{\uparrow}\hat J_{+}\hat S_{-} g_0^{\downarrow}, \quad
\hat G_0^{-} = \hat G_0^{\downarrow}\hat J_{-}\hat S_{+} g_0^{\uparrow}, 
\label{G_0}
\end{eqnarray}
and
\begin{eqnarray}
\hat D_0^{\uparrow} &=& \left[ 1 - \hat G^{\uparrow}
\left(\hat w^{\uparrow} - \hat \Sigma^{\uparrow} +
\hat J_{+}\hat n_{+}\hat g_0^{\downarrow}\hat J_{-} \right)
\right]^{-1}, \nonumber\\
\hat D_0^{\downarrow} &=& \left[ 1 - \hat G^{\downarrow}
\left(\hat w^{\downarrow} - \hat \Sigma^{\downarrow} +
\hat J_{-}\hat n_{-}\hat g_0^{\uparrow}\hat J_{+} \right)
\right]^{-1}, \nonumber \\
\hat D_0^{+} &=& \hat G_0^{\uparrow}\hat J_{+} \hat S_{-} d_0^{\downarrow}, \quad
\hat D_0^{-} = \hat G_0^{\downarrow}\hat J_{-} \hat S_{+} d_0^{\uparrow}, 
\label{D_0}
\end{eqnarray}
where "denominators" $d_0^{\uparrow(\downarrow)}$ are given by
\[
\hat d_0^{\uparrow(\downarrow)} = \left[ 1 - \hat G^{\uparrow(\downarrow)}
\left(\hat w^{\uparrow(\downarrow)} -
\hat \Sigma^{\uparrow(\downarrow)}\right)\right]^{-1}.
\]
Now, using~(\ref{G_0}) and (\ref{D_0}) and taking into account the
obvious relation $\hat t_{00}=(\hat W -\hat\Sigma)\hat D_0$, one 
finds
\begin{eqnarray}
\hat t_0^{\uparrow} & = & \left(\hat w^{\uparrow} - \hat\Sigma^{\uparrow} +
\hat J_{+}\hat n_{+}\hat g_0^{\downarrow}\hat J_{-}\right)\hat D_0^{\uparrow},
\nonumber \\
\hat t_0^{\downarrow} & = & \left(\hat w^{\downarrow} - \hat\Sigma^{\downarrow} +
\hat J_{-}\hat n_{-}\hat g_0^{\uparrow}\hat J_{+}\right)\hat D_0^{\downarrow},
\nonumber\\
\label{t_00}
\hat t_0^{+} & = & \hat d_0^{\uparrow T}\hat J_{+}\hat D_0^{\downarrow},\quad
\hat t_0^{-} = \hat d_0^{\downarrow T}\hat J_{-}\hat D_0^{\uparrow}.
\end{eqnarray}
According to the ASF, $\hat G_0^{\uparrow(\downarrow)}$ and
$\hat t_0^{\uparrow(\downarrow)}$ represent the configurationally
averaged quantities, and after the averaging over
magnon degrees of freedom one must get
$\langle\hat G_0^{\uparrow(\downarrow)}\rangle_b =
\hat G^{\uparrow(\downarrow)}$ and
$\langle\hat t_0^{\uparrow(\downarrow)}\rangle_b = 0$.
It means that due to the CPA self-consistency conditions the 
averaged scattering $t$-matrix must vanish and averaged electron propagator 
has to be equal to the effective Green's function~(\ref{G_int}), determined by 
the self-energy operator $\hat\Sigma$. Carrying out the averaging procedure over 
magnon degrees of freedom the similar way as it was proposed earlier
in Sec.~IIC, it is 
possible to satisfy both of these condition if one assumes that
\begin{equation}
\begin{array}{ccc}
\hat\Sigma^{\uparrow} &=& \hat\omega^{\uparrow}(n)
+ \hat J_{+}(n)n\hat g_0^{\downarrow}(n)\hat J_{-}(n), \smallskip \\
\hat\Sigma^{\downarrow} &=& \hat\omega^{\downarrow}(n) +
\hat J_{-}(n)n\hat g_0^{\uparrow}(n)\hat J_{+}(n).
\end{array}
\label{CPA_iter}
\end{equation}
We have pointed out the explicit dependence of matrices
$\hat\omega^{\uparrow(\downarrow)}(n)$,
$\hat J_{\pm}(n)$ and
$\hat g_0^{\uparrow(\downarrow)}(n)$ on the average magnon number
$n$ (Eq.~\ref{n_T}). This dependence is assumed to be the same
as it comes from the initial definition of operators
$\hat\omega^{\uparrow(\downarrow)}$, $\hat J_{\pm}$, and
$\hat g_0^{\uparrow(\downarrow)}$ as functions on $\hat n_{\pm}$
[Eqs.\ (\ref{w_up_down})--(\ref{g0_up_down})],
that is the consequence of the adopted approximate averaging
procedure. The matrices $\hat\omega^{\uparrow(\downarrow)}$,
$\hat J_{\pm}$ and $\hat g_0^{\uparrow(\downarrow)}$ functionally depend on
$\Sigma^{\uparrow(\downarrow)}$. Due to that, the system~(\ref{CPA_iter})
represents the alternative to Eq.~(\ref{CPA_system}) form of the CPA conditions
and it can be simply solved by means of successive numerical
iterations.

\end{document}